% File jytex.tex, for jyTeX version 2.6M (June 1992)
% Copyright (c) 1991, 1992 by Jonathan P. Yamron
% For full documentation, "get jydoc" from hep-ph@xxx.lanl.gov
%   Problems?  Contact brahm@theory3.caltech.edu.

\catcode`\@=11

%*****************************************************************************

\message{Loading jyTeX fonts...}

%************************************************************
%*
%*             Available fonts
%*
%************************************************************

%************** 5-point fonts *******************************

\font\vptrm=cmr5 \font\vptmit=cmmi5 \font\vptsy=cmsy5 \font\vptbf=cmbx5

\skewchar\vptmit='177 \skewchar\vptsy='60 \fontdimen16
\vptsy=\the\fontdimen17 \vptsy

\def\vpt{\ifmmode\err@badsizechange\else
     \@mathfontinit
     \textfont0=\vptrm  \scriptfont0=\vptrm  \scriptscriptfont0=\vptrm
     \textfont1=\vptmit \scriptfont1=\vptmit \scriptscriptfont1=\vptmit
     \textfont2=\vptsy  \scriptfont2=\vptsy  \scriptscriptfont2=\vptsy
     \textfont3=\xptex  \scriptfont3=\xptex  \scriptscriptfont3=\xptex
     \textfont\bffam=\vptbf
     \scriptfont\bffam=\vptbf
     \scriptscriptfont\bffam=\vptbf
     \@fontstyleinit
     \def\rm{\vptrm\fam=\z@}%
     \def\bf{\vptbf\fam=\bffam}%
     \def\oldstyle{\vptmit\fam=\@ne}%
     \rm\fi}

%************** 6-point fonts *******************************

\font\viptrm=cmr6 \font\viptmit=cmmi6 \font\viptsy=cmsy6
\font\viptbf=cmbx6

\skewchar\viptmit='177 \skewchar\viptsy='60 \fontdimen16
\viptsy=\the\fontdimen17 \viptsy

\def\vipt{\ifmmode\err@badsizechange\else
     \@mathfontinit
     \textfont0=\viptrm  \scriptfont0=\vptrm  \scriptscriptfont0=\vptrm
     \textfont1=\viptmit \scriptfont1=\vptmit \scriptscriptfont1=\vptmit
     \textfont2=\viptsy  \scriptfont2=\vptsy  \scriptscriptfont2=\vptsy
     \textfont3=\xptex   \scriptfont3=\xptex  \scriptscriptfont3=\xptex
     \textfont\bffam=\viptbf
     \scriptfont\bffam=\vptbf
     \scriptscriptfont\bffam=\vptbf
     \@fontstyleinit
     \def\rm{\viptrm\fam=\z@}%
     \def\bf{\viptbf\fam=\bffam}%
     \def\oldstyle{\viptmit\fam=\@ne}%
     \rm\fi}
%************** 7-point fonts *******************************

\font\viiptrm=cmr7 \font\viiptmit=cmmi7 \font\viiptsy=cmsy7
\font\viiptit=cmti7 \font\viiptbf=cmbx7

\skewchar\viiptmit='177 \skewchar\viiptsy='60 \fontdimen16
\viiptsy=\the\fontdimen17 \viiptsy

\def\viipt{\ifmmode\err@badsizechange\else
     \@mathfontinit
     \textfont0=\viiptrm  \scriptfont0=\vptrm  \scriptscriptfont0=\vptrm
     \textfont1=\viiptmit \scriptfont1=\vptmit \scriptscriptfont1=\vptmit
     \textfont2=\viiptsy  \scriptfont2=\vptsy  \scriptscriptfont2=\vptsy
     \textfont3=\xptex    \scriptfont3=\xptex  \scriptscriptfont3=\xptex
     \textfont\itfam=\viiptit
     \scriptfont\itfam=\viiptit
     \scriptscriptfont\itfam=\viiptit
     \textfont\bffam=\viiptbf
     \scriptfont\bffam=\vptbf
     \scriptscriptfont\bffam=\vptbf
     \@fontstyleinit
     \def\rm{\viiptrm\fam=\z@}%
     \def\it{\viiptit\fam=\itfam}%
     \def\bf{\viiptbf\fam=\bffam}%
     \def\oldstyle{\viiptmit\fam=\@ne}%
     \rm\fi}

%************** 8-point fonts *******************************

\font\viiiptrm=cmr8 \font\viiiptmit=cmmi8 \font\viiiptsy=cmsy8
\font\viiiptit=cmti8
%\font\viiiptsl=cmsl8
\font\viiiptbf=cmbx8
%\font\viiipttt=cmtt8
%\font\viiiptss=cmss8

\skewchar\viiiptmit='177 \skewchar\viiiptsy='60 \fontdimen16
\viiiptsy=\the\fontdimen17 \viiiptsy

\def\viiipt{\ifmmode\err@badsizechange\else
     \@mathfontinit
     \textfont0=\viiiptrm  \scriptfont0=\viptrm  \scriptscriptfont0=\vptrm
     \textfont1=\viiiptmit \scriptfont1=\viptmit \scriptscriptfont1=\vptmit
     \textfont2=\viiiptsy  \scriptfont2=\viptsy  \scriptscriptfont2=\vptsy
     \textfont3=\xptex     \scriptfont3=\xptex   \scriptscriptfont3=\xptex
     \textfont\itfam=\viiiptit
     \scriptfont\itfam=\viiptit
     \scriptscriptfont\itfam=\viiptit
     \textfont\bffam=\viiiptbf
     \scriptfont\bffam=\viptbf
     \scriptscriptfont\bffam=\vptbf
     \@fontstyleinit
     \def\rm{\viiiptrm\fam=\z@}%
     \def\it{\viiiptit\fam=\itfam}%
     \def\bf{\viiiptbf\fam=\bffam}%
     \def\oldstyle{\viiiptmit\fam=\@ne}%
     \rm\fi}

%************** Optional 9-point fonts **********************

\def\getixpt{%
     \font\ixptrm=cmr9
     \font\ixptmit=cmmi9
     \font\ixptsy=cmsy9
     \font\ixptit=cmti9
%     \font\ixptsl=cmsl9
     \font\ixptbf=cmbx9
%     \font\ixpttt=cmtt9
%     \font\ixptss=cmss9
     \skewchar\ixptmit='177 \skewchar\ixptsy='60
     \fontdimen16 \ixptsy=\the\fontdimen17 \ixptsy}

\def\ixpt{\ifmmode\err@badsizechange\else
     \@mathfontinit
     \textfont0=\ixptrm  \scriptfont0=\viiptrm  \scriptscriptfont0=\vptrm
     \textfont1=\ixptmit \scriptfont1=\viiptmit \scriptscriptfont1=\vptmit
     \textfont2=\ixptsy  \scriptfont2=\viiptsy  \scriptscriptfont2=\vptsy
     \textfont3=\xptex   \scriptfont3=\xptex    \scriptscriptfont3=\xptex
     \textfont\itfam=\ixptit
     \scriptfont\itfam=\viiptit
     \scriptscriptfont\itfam=\viiptit
     \textfont\bffam=\ixptbf
     \scriptfont\bffam=\viiptbf
     \scriptscriptfont\bffam=\vptbf
     \@fontstyleinit
     \def\rm{\ixptrm\fam=\z@}%
     \def\it{\ixptit\fam=\itfam}%
     \def\bf{\ixptbf\fam=\bffam}%
     \def\oldstyle{\ixptmit\fam=\@ne}%
     \rm\fi}

%************** 10-point fonts ******************************

\font\xptrm=cmr10 \font\xptmit=cmmi10 \font\xptsy=cmsy10
\font\xptex=cmex10 \font\xptit=cmti10 \font\xptsl=cmsl10
\font\xptbf=cmbx10 \font\xpttt=cmtt10 \font\xptss=cmss10
\font\xptsc=cmcsc10 \font\xptbfs=cmb10 \font\xptbmit=cmmib10

\skewchar\xptmit='177 \skewchar\xptbmit='177 \skewchar\xptsy='60
\fontdimen16 \xptsy=\the\fontdimen17 \xptsy

\def\xpt{\ifmmode\err@badsizechange\else
     \@mathfontinit
     \textfont0=\xptrm  \scriptfont0=\viiptrm  \scriptscriptfont0=\vptrm
     \textfont1=\xptmit \scriptfont1=\viiptmit \scriptscriptfont1=\vptmit
     \textfont2=\xptsy  \scriptfont2=\viiptsy  \scriptscriptfont2=\vptsy
     \textfont3=\xptex  \scriptfont3=\xptex    \scriptscriptfont3=\xptex
     \textfont\itfam=\xptit
     \scriptfont\itfam=\viiptit
     \scriptscriptfont\itfam=\viiptit
     \textfont\bffam=\xptbf
     \scriptfont\bffam=\viiptbf
     \scriptscriptfont\bffam=\vptbf
     \textfont\bfsfam=\xptbfs
     \scriptfont\bfsfam=\viiptbf
     \scriptscriptfont\bfsfam=\vptbf
     \textfont\bmitfam=\xptbmit
     \scriptfont\bmitfam=\viiptmit
     \scriptscriptfont\bmitfam=\vptmit
     \@fontstyleinit
     \def\rm{\xptrm\fam=\z@}%
     \def\it{\xptit\fam=\itfam}%
     \def\sl{\xptsl}%
     \def\bf{\xptbf\fam=\bffam}%
     \def\tt{\xpttt}%
     \def\ss{\xptss}%
     \def\sc{\xptsc}%
     \def\bfs{\xptbfs\fam=\bfsfam}%
     \def\bmit{\fam=\bmitfam}%
     \def\oldstyle{\xptmit\fam=\@ne}%
     \rm\fi}

%************** Optional 11-point fonts *********************

\def\getxipt{%
     \font\xiptrm=cmr10  scaled\magstephalf
     \font\xiptmit=cmmi10 scaled\magstephalf
     \font\xiptsy=cmsy10 scaled\magstephalf
     \font\xiptex=cmex10 scaled\magstephalf
     \font\xiptit=cmti10 scaled\magstephalf
     \font\xiptsl=cmsl10 scaled\magstephalf
     \font\xiptbf=cmbx10 scaled\magstephalf
     \font\xipttt=cmtt10 scaled\magstephalf
     \font\xiptss=cmss10 scaled\magstephalf
     \skewchar\xiptmit='177 \skewchar\xiptsy='60
     \fontdimen16 \xiptsy=\the\fontdimen17 \xiptsy}

\def\xipt{\ifmmode\err@badsizechange\else
     \@mathfontinit
     \textfont0=\xiptrm  \scriptfont0=\viiiptrm  \scriptscriptfont0=\viptrm
     \textfont1=\xiptmit \scriptfont1=\viiiptmit \scriptscriptfont1=\viptmit
     \textfont2=\xiptsy  \scriptfont2=\viiiptsy  \scriptscriptfont2=\viptsy
     \textfont3=\xiptex  \scriptfont3=\xptex     \scriptscriptfont3=\xptex
     \textfont\itfam=\xiptit
     \scriptfont\itfam=\viiiptit
     \scriptscriptfont\itfam=\viiptit
     \textfont\bffam=\xiptbf
     \scriptfont\bffam=\viiiptbf
     \scriptscriptfont\bffam=\viptbf
     \@fontstyleinit
     \def\rm{\xiptrm\fam=\z@}%
     \def\it{\xiptit\fam=\itfam}%
     \def\sl{\xiptsl}%
     \def\bf{\xiptbf\fam=\bffam}%
     \def\tt{\xipttt}%
     \def\ss{\xiptss}%
     \def\oldstyle{\xiptmit\fam=\@ne}%
     \rm\fi}

%************** 12-point fonts ******************************

\font\xiiptrm=cmr12 \font\xiiptmit=cmmi12 \font\xiiptsy=cmsy10
scaled\magstep1 \font\xiiptex=cmex10  scaled\magstep1
\font\xiiptit=cmti12 \font\xiiptsl=cmsl12 \font\xiiptbf=cmbx12
%\font\xiipttt=cmtt12
\font\xiiptss=cmss12 \font\xiiptsc=cmcsc10 scaled\magstep1
\font\xiiptbfs=cmb10  scaled\magstep1 \font\xiiptbmit=cmmib10
scaled\magstep1

\skewchar\xiiptmit='177 \skewchar\xiiptbmit='177 \skewchar\xiiptsy='60
\fontdimen16 \xiiptsy=\the\fontdimen17 \xiiptsy

\def\xiipt{\ifmmode\err@badsizechange\else
     \@mathfontinit
     \textfont0=\xiiptrm  \scriptfont0=\viiiptrm  \scriptscriptfont0=\viptrm
     \textfont1=\xiiptmit \scriptfont1=\viiiptmit \scriptscriptfont1=\viptmit
     \textfont2=\xiiptsy  \scriptfont2=\viiiptsy  \scriptscriptfont2=\viptsy
     \textfont3=\xiiptex  \scriptfont3=\xptex     \scriptscriptfont3=\xptex
     \textfont\itfam=\xiiptit
     \scriptfont\itfam=\viiiptit
     \scriptscriptfont\itfam=\viiptit
     \textfont\bffam=\xiiptbf
     \scriptfont\bffam=\viiiptbf
     \scriptscriptfont\bffam=\viptbf
     \textfont\bfsfam=\xiiptbfs
     \scriptfont\bfsfam=\viiiptbf
     \scriptscriptfont\bfsfam=\viptbf
     \textfont\bmitfam=\xiiptbmit
     \scriptfont\bmitfam=\viiiptmit
     \scriptscriptfont\bmitfam=\viptmit
     \@fontstyleinit
     \def\rm{\xiiptrm\fam=\z@}%
     \def\it{\xiiptit\fam=\itfam}%
     \def\sl{\xiiptsl}%
     \def\bf{\xiiptbf\fam=\bffam}%
     \def\tt{\xiipttt}%
     \def\ss{\xiiptss}%
     \def\sc{\xiiptsc}%
     \def\bfs{\xiiptbfs\fam=\bfsfam}%
     \def\bmit{\fam=\bmitfam}%
     \def\oldstyle{\xiiptmit\fam=\@ne}%
     \rm\fi}

%************** Optional 13-point fonts *********************

\def\getxiiipt{%
     \font\xiiiptrm=cmr12  scaled\magstephalf
     \font\xiiiptmit=cmmi12 scaled\magstephalf
     \font\xiiiptsy=cmsy9  scaled\magstep2
     \font\xiiiptit=cmti12 scaled\magstephalf
     \font\xiiiptsl=cmsl12 scaled\magstephalf
     \font\xiiiptbf=cmbx12 scaled\magstephalf
     \font\xiiipttt=cmtt12 scaled\magstephalf
     \font\xiiiptss=cmss12 scaled\magstephalf
     \skewchar\xiiiptmit='177 \skewchar\xiiiptsy='60
     \fontdimen16 \xiiiptsy=\the\fontdimen17 \xiiiptsy}

\def\xiiipt{\ifmmode\err@badsizechange\else
     \@mathfontinit
     \textfont0=\xiiiptrm  \scriptfont0=\xptrm  \scriptscriptfont0=\viiptrm
     \textfont1=\xiiiptmit \scriptfont1=\xptmit \scriptscriptfont1=\viiptmit
     \textfont2=\xiiiptsy  \scriptfont2=\xptsy  \scriptscriptfont2=\viiptsy
     \textfont3=\xivptex   \scriptfont3=\xptex  \scriptscriptfont3=\xptex
     \textfont\itfam=\xiiiptit
     \scriptfont\itfam=\xptit
     \scriptscriptfont\itfam=\viiptit
     \textfont\bffam=\xiiiptbf
     \scriptfont\bffam=\xptbf
     \scriptscriptfont\bffam=\viiptbf
     \@fontstyleinit
     \def\rm{\xiiiptrm\fam=\z@}%
     \def\it{\xiiiptit\fam=\itfam}%
     \def\sl{\xiiiptsl}%
     \def\bf{\xiiiptbf\fam=\bffam}%
     \def\tt{\xiiipttt}%
     \def\ss{\xiiiptss}%
     \def\oldstyle{\xiiiptmit\fam=\@ne}%
     \rm\fi}

%************** 14-point fonts ******************************

\font\xivptrm=cmr12   scaled\magstep1 \font\xivptmit=cmmi12
scaled\magstep1 \font\xivptsy=cmsy10  scaled\magstep2
\font\xivptex=cmex10  scaled\magstep2 \font\xivptit=cmti12
scaled\magstep1 \font\xivptsl=cmsl12  scaled\magstep1
\font\xivptbf=cmbx12  scaled\magstep1
%\font\xivpttt=cmtt12  scaled\magstep1
\font\xivptss=cmss12  scaled\magstep1 \font\xivptsc=cmcsc10
scaled\magstep2 \font\xivptbfs=cmb10  scaled\magstep2
\font\xivptbmit=cmmib10 scaled\magstep2

\skewchar\xivptmit='177 \skewchar\xivptbmit='177 \skewchar\xivptsy='60
\fontdimen16 \xivptsy=\the\fontdimen17 \xivptsy

\def\xivpt{\ifmmode\err@badsizechange\else
     \@mathfontinit
     \textfont0=\xivptrm  \scriptfont0=\xptrm  \scriptscriptfont0=\viiptrm
     \textfont1=\xivptmit \scriptfont1=\xptmit \scriptscriptfont1=\viiptmit
     \textfont2=\xivptsy  \scriptfont2=\xptsy  \scriptscriptfont2=\viiptsy
     \textfont3=\xivptex  \scriptfont3=\xptex  \scriptscriptfont3=\xptex
     \textfont\itfam=\xivptit
     \scriptfont\itfam=\xptit
     \scriptscriptfont\itfam=\viiptit
     \textfont\bffam=\xivptbf
     \scriptfont\bffam=\xptbf
     \scriptscriptfont\bffam=\viiptbf
     \textfont\bfsfam=\xivptbfs
     \scriptfont\bfsfam=\xptbfs
     \scriptscriptfont\bfsfam=\viiptbf
     \textfont\bmitfam=\xivptbmit
     \scriptfont\bmitfam=\xptbmit
     \scriptscriptfont\bmitfam=\viiptmit
     \@fontstyleinit
     \def\rm{\xivptrm\fam=\z@}%
     \def\it{\xivptit\fam=\itfam}%
     \def\sl{\xivptsl}%
     \def\bf{\xivptbf\fam=\bffam}%
     \def\tt{\xivpttt}%
     \def\ss{\xivptss}%
     \def\sc{\xivptsc}%
     \def\bfs{\xivptbfs\fam=\bfsfam}%
     \def\bmit{\fam=\bmitfam}%
     \def\oldstyle{\xivptmit\fam=\@ne}%
     \rm\fi}

%************** 17-point fonts ******************************

\font\xviiptrm=cmr17 \font\xviiptmit=cmmi12 scaled\magstep2
\font\xviiptsy=cmsy10 scaled\magstep3 \font\xviiptex=cmex10
scaled\magstep3 \font\xviiptit=cmti12 scaled\magstep2
\font\xviiptbf=cmbx12 scaled\magstep2 \font\xviiptbfs=cmb10
scaled\magstep3

\skewchar\xviiptmit='177 \skewchar\xviiptsy='60 \fontdimen16
\xviiptsy=\the\fontdimen17 \xviiptsy

\def\xviipt{\ifmmode\err@badsizechange\else
     \@mathfontinit
     \textfont0=\xviiptrm  \scriptfont0=\xiiptrm  \scriptscriptfont0=\viiiptrm
     \textfont1=\xviiptmit \scriptfont1=\xiiptmit \scriptscriptfont1=\viiiptmit
     \textfont2=\xviiptsy  \scriptfont2=\xiiptsy  \scriptscriptfont2=\viiiptsy
     \textfont3=\xviiptex  \scriptfont3=\xiiptex  \scriptscriptfont3=\xptex
     \textfont\itfam=\xviiptit
     \scriptfont\itfam=\xiiptit
     \scriptscriptfont\itfam=\viiiptit
     \textfont\bffam=\xviiptbf
     \scriptfont\bffam=\xiiptbf
     \scriptscriptfont\bffam=\viiiptbf
     \textfont\bfsfam=\xviiptbfs
     \scriptfont\bfsfam=\xiiptbfs
     \scriptscriptfont\bfsfam=\viiiptbf
     \@fontstyleinit
     \def\rm{\xviiptrm\fam=\z@}%
     \def\it{\xviiptit\fam=\itfam}%
     \def\bf{\xviiptbf\fam=\bffam}%
     \def\bfs{\xviiptbfs\fam=\bfsfam}%
     \def\oldstyle{\xviiptmit\fam=\@ne}%
     \rm\fi}

%************** 21-point fonts ******************************

\font\xxiptrm=cmr17  scaled\magstep1
%\font\xxiptmit=cmmi12 scaled\magstep3
%\font\xxiptsy=cmsy10 scaled\magstep4
%\font\xxiptex=cmex10 scaled\magstep4
%\font\xxiptbf=cmbx12 scaled\magstep3

%\skewchar\xxiptmit='177 \skewchar\xxiptsy='60
%\fontdimen16 \xxiptsy=\the\fontdimen17 \xxiptsy

\def\xxipt{\ifmmode\err@badsizechange\else
     \@mathfontinit
%     \textfont0=\xxiptrm  \scriptfont0=\xivptrm  \scriptscriptfont0=\xptrm
%     \textfont1=\xxiptmit \scriptfont1=\xivptmit \scriptscriptfont1=\xptmit
%     \textfont2=\xxiptsy  \scriptfont2=\xivptsy  \scriptscriptfont2=\xptsy
%     \textfont3=\xxiptex  \scriptfont3=\xivptex  \scriptscriptfont3=\xptex
%     \textfont\bffam=\xxiptbf
%     \scriptfont\bffam=\xivptbf
%     \scriptscriptfont\bffam=\xptbf
     \@fontstyleinit
     \def\rm{\xxiptrm\fam=\z@}%
     \rm\fi}

%************** 25-point fonts ******************************

\font\xxvptrm=cmr17  scaled\magstep2
%\font\xxvptmit=cmmi12 scaled\magstep4
%\font\xxvptsy=cmsy10 scaled\magstep5
%\font\xxvptex=cmex10 scaled\magstep5
%\font\xxvptbf=cmbx12 scaled\magstep4

%\skewchar\xxvptmit='177 \skewchar\xxvptsy='60
%\fontdimen16 \xxvptsy=\the\fontdimen17 \xxvptsy

\def\xxvpt{\ifmmode\err@badsizechange\else
     \@mathfontinit
%     \textfont0=\xxvptrm  \scriptfont0=\xviiptrm  \scriptscriptfont0=\xiiptrm
%     \textfont1=\xxvptmit \scriptfont1=\xviiptmit \scriptscriptfont1=\xiiptmit
%     \textfont2=\xxvptsy  \scriptfont2=\xviiptsy  \scriptscriptfont2=\xiiptsy
%     \textfont3=\xxvptex  \scriptfont3=\xviiptex  \scriptscriptfont3=\xiiptex
%     \textfont\bffam=\xxvptbf
%     \scriptfont\bffam=\xviiptbf
%     \scriptscriptfont\bffam=\xiiptbf
     \@fontstyleinit
     \def\rm{\xxvptrm\fam=\z@}%
     \rm\fi}

%************** Other fonts *********************************

%\font\dummy=dummy

%******************************************************************************

\message{Loading jyTeX macros...}

%************************************************************
%*
%*              Simple modifications to plain
%*
%************************************************************
\message{modifications to plain.tex,}

% The "\outer" qualifier is removed from the definitions of \newcount through
% \newif so that they may be used in definitions.  \newif is also changed to
% make \if commands globally defined.

\def\newcount{\alloc@0\count\countdef\insc@unt}
\def\newdimen{\alloc@1\dimen\dimendef\insc@unt}
\def\newskip{\alloc@2\skip\skipdef\insc@unt}
\def\newmuskip{\alloc@3\muskip\muskipdef\@cclvi}
\def\newbox{\alloc@4\box\chardef\insc@unt}
\def\newtoks{\alloc@5\toks\toksdef\@cclvi}
\def\newhelp#1#2{\newtoks#1\global#1\expandafter{\csname#2\endcsname}}
\def\newread{\alloc@6\read\chardef\sixt@@n}
\def\newwrite{\alloc@7\write\chardef\sixt@@n}
\def\newfam{\alloc@8\fam\chardef\sixt@@n}
\def\newinsert#1{\global\advance\insc@unt by\m@ne
     \ch@ck0\insc@unt\count
     \ch@ck1\insc@unt\dimen
     \ch@ck2\insc@unt\skip
     \ch@ck4\insc@unt\box
     \allocationnumber=\insc@unt
     \global\chardef#1=\allocationnumber
     \wlog{\string#1=\string\insert\the\allocationnumber}}
\def\newif#1{\count@\escapechar \escapechar\m@ne
     \expandafter\expandafter\expandafter
          \xdef\@if#1{true}{\let\noexpand#1=\noexpand\iftrue}%
     \expandafter\expandafter\expandafter
          \xdef\@if#1{false}{\let\noexpand#1=\noexpand\iffalse}%
     \global\@if#1{false}\escapechar=\count@}

%************** Some parameter changes **********************

\newlinechar=`\^^J
\overfullrule=0pt

%************** Font-related modifications ******************

% The plain fonts are mapped onto the corresponding jyTeX fonts

% Some control sequences are disabled.

\let\itfam=\undefined

\let\bffam=\undefined

\count18=3

% German sharp s is given a new name (\ss is already taken)

\chardef\sharps="19

% The mathcode assignments of characters in the math italic font are changed to
% allow for switching to boldface.

\mathchardef\alpha="710B \mathchardef\beta="710C \mathchardef\gamma="710D
\mathchardef\delta="710E \mathchardef\epsilon="710F
\mathchardef\zeta="7110 \mathchardef\eta="7111 \mathchardef\theta="7112
\mathchardef\iota="7113 \mathchardef\kappa="7114
\mathchardef\lambda="7115 \mathchardef\mu="7116 \mathchardef\nu="7117
\mathchardef\xi="7118 \mathchardef\pi="7119 \mathchardef\rho="711A
\mathchardef\sigma="711B \mathchardef\tau="711C
\mathchardef\upsilon="711D \mathchardef\phi="711E \mathchardef\chi="711F
\mathchardef\psi="7120 \mathchardef\omega="7121
\mathchardef\varepsilon="7122 \mathchardef\vartheta="7123
\mathchardef\varpi="7124 \mathchardef\varrho="7125
\mathchardef\varsigma="7126 \mathchardef\varphi="7127
\mathchardef\imath="717B \mathchardef\jmath="717C \mathchardef\ell="7160
\mathchardef\wp="717D \mathchardef\partial="7140 \mathchardef\flat="715B
\mathchardef\natural="715C \mathchardef\sharp="715D

%************** Miscellaneous changes ***********************

% The dimension \p@ (1pt) is replaced with \rp@ (relative pt, defined below),
% whose size is determined by the base type size of the document.

\def\angle{{\vbox{\ialign{$\m@th\scriptstyle##$\crcr
     \not\mathrel{\mkern14mu}\crcr
     \noalign{\nointerlineskip}
     \mkern2.5mu\leaders\hrule height.34\rp@\hfill\mkern2.5mu\crcr}}}}
\def\vdots{\vbox{\baselineskip4\rp@ \lineskiplimit\z@
     \kern6\rp@\hbox{.}\hbox{.}\hbox{.}}}
\def\ddots{\mathinner{\mkern1mu\raise7\rp@\vbox{\kern7\rp@\hbox{.}}\mkern2mu
     \raise4\rp@\hbox{.}\mkern2mu\raise\rp@\hbox{.}\mkern1mu}}
\def\overrightarrow#1{\vbox{\ialign{##\crcr
     \rightarrowfill\crcr
     \noalign{\kern-\rp@\nointerlineskip}
     $\hfil\displaystyle{#1}\hfil$\crcr}}}
\def\overleftarrow#1{\vbox{\ialign{##\crcr
     \leftarrowfill\crcr
     \noalign{\kern-\rp@\nointerlineskip}
     $\hfil\displaystyle{#1}\hfil$\crcr}}}
\def\overbrace#1{\mathop{\vbox{\ialign{##\crcr
     \noalign{\kern3\rp@}
     \downbracefill\crcr
     \noalign{\kern3\rp@\nointerlineskip}
     $\hfil\displaystyle{#1}\hfil$\crcr}}}\limits}
\def\underbrace#1{\mathop{\vtop{\ialign{##\crcr
     $\hfil\displaystyle{#1}\hfil$\crcr
     \noalign{\kern3\rp@\nointerlineskip}
     \upbracefill\crcr
     \noalign{\kern3\rp@}}}}\limits}
\def\big#1{{\hbox{$\left#1\vbox to8.5\rp@ {}\right.\n@space$}}}
\def\Big#1{{\hbox{$\left#1\vbox to11.5\rp@ {}\right.\n@space$}}}
\def\bigg#1{{\hbox{$\left#1\vbox to14.5\rp@ {}\right.\n@space$}}}
\def\Bigg#1{{\hbox{$\left#1\vbox to17.5\rp@ {}\right.\n@space$}}}
\def\@vereq#1#2{\lower.5\rp@\vbox{\baselineskip\z@skip\lineskip-.5\rp@
     \ialign{$\m@th#1\hfil##\hfil$\crcr#2\crcr=\crcr}}}
\def\rlh@#1{\vcenter{\hbox{\ooalign{\raise2\rp@
     \hbox{$#1\rightharpoonup$}\crcr
     $#1\leftharpoondown$}}}}
\def\bordermatrix#1{\begingroup\m@th
     \setbox\z@\vbox{%
          \def\cr{\crcr\noalign{\kern2\rp@\global\let\cr\endline}}%
          \ialign{$##$\hfil\kern2\rp@\kern\p@renwd
               &\thinspace\hfil$##$\hfil&&\quad\hfil$##$\hfil\crcr
               \omit\strut\hfil\crcr
               \noalign{\kern-\baselineskip}%
               #1\crcr\omit\strut\cr}}%
     \setbox\tw@\vbox{\unvcopy\z@\global\setbox\@ne\lastbox}%
     \setbox\tw@\hbox{\unhbox\@ne\unskip\global\setbox\@ne\lastbox}%
     \setbox\tw@\hbox{$\kern\wd\@ne\kern-\p@renwd\left(\kern-\wd\@ne
          \global\setbox\@ne\vbox{\box\@ne\kern2\rp@}%
          \vcenter{\kern-\ht\@ne\unvbox\z@\kern-\baselineskip}%
          \,\right)$}%
     \null\;\vbox{\kern\ht\@ne\box\tw@}\endgroup}
\def\endinsert{\egroup
     \if@mid\dimen@\ht\z@
          \advance\dimen@\dp\z@
          \advance\dimen@12\rp@
          \advance\dimen@\pagetotal
          \ifdim\dimen@>\pagegoal\@midfalse\p@gefalse\fi
     \fi
     \if@mid\bigskip\box\z@
          \bigbreak
     \else\insert\topins{\penalty100 \splittopskip\z@skip
               \splitmaxdepth\maxdimen\floatingpenalty\z@
               \ifp@ge\dimen@\dp\z@
                    \vbox to\vsize{\unvbox\z@\kern-\dimen@}%
               \else\box\z@\nobreak\bigskip
               \fi}%
     \fi
     \endgroup}

% \normalbaselines is removed from \cases and \matrix.

\def\cases#1{\left\{\,\vcenter{\m@th
     \ialign{$##\hfil$&\quad##\hfil\crcr#1\crcr}}\right.}
\def\matrix#1{\null\,\vcenter{\m@th
     \ialign{\hfil$##$\hfil&&\quad\hfil$##$\hfil\crcr
          \mathstrut\crcr
          \noalign{\kern-\baselineskip}
          #1\crcr
          \mathstrut\crcr
          \noalign{\kern-\baselineskip}}}\,}

% \raggedbottom modified slightly

\newif\ifraggedbottom

\def\raggedbottom{\ifraggedbottom\else
     \advance\topskip by\z@ plus60pt \raggedbottomtrue\fi}%
\def\normalbottom{\ifraggedbottom
     \advance\topskip by\z@ plus-60pt \raggedbottomfalse\fi}

%************************************************************
%*
%*              Miscellaneous definitions
%*
%************************************************************
\message{hacks,}

%************** Hack registers ******************************

\toksdef\toks@i=1 \toksdef\toks@ii=2

%************** Basic macros ********************************

\def\TeX{T\kern-.1667em \lower.5ex \hbox{E}\kern-.125em X\null}
\def\jyTeX{{\leavevmode
     \raise.587ex \hbox{\it\j}\kern-.1em \lower.048ex \hbox{\it y}\kern-.12em
     \TeX}}

\let\then=\iftrue
\def\ifnoarg#1\then{\def\hack@{#1}\ifx\hack@\empty}
\def\ifundefined#1\then{%
     \expandafter\ifx\csname\expandafter\blank\string#1\endcsname\relax}
\def\useif#1\then{\csname#1\endcsname}
\def\usename#1{\csname#1\endcsname}
\def\useafter#1#2{\expandafter#1\csname#2\endcsname}

% Modify so that I can have \loop's within \loop's?
\long\def\loop#1\repeat{\def\@iterate{#1\expandafter\@iterate\fi}\@iterate
     \let\@iterate=\relax}
%\long\def\loop#1\repeat{\def\@loopbody{#1}\@iterate}
%\def\@iterate{\@loopbody\let\next=\@iterate\else\let\next=\relax\fi\next}

\let\TeXend=\end
\def\begin#1{\begingroup\def\@@blockname{#1}\usename{begin#1}}
\def\end#1{\usename{end#1}\def\hack@{#1}%
     \ifx\@@blockname\hack@
          \endgroup
     \else\err@badgroup\hack@\@@blockname
     \fi}
\def\@@blockname{}

\def\defaultoption[#1]#2{%
     \def\hack@{\ifx\hack@ii[\toks@={#2}\else\toks@={#2[#1]}\fi\the\toks@}%
     \futurelet\hack@ii\hack@}

\def\markup#1{\let\@@marksf=\empty
     \ifhmode\edef\@@marksf{\spacefactor=\the\spacefactor\relax}\/\fi
     ${}^{\hbox{\subscriptfonts#1}}$\@@marksf}

%************** Time registers ******************************

\newtoks\shortyear
\newtoks\militaryhour
\newtoks\standardhour
\newtoks\minute
\newtoks\amorpm

\def\settime{\count@=\time\divide\count@ by60
     \militaryhour=\expandafter{\number\count@}%
     {\multiply\count@ by-60 \advance\count@ by\time
          \xdef\hack@{\ifnum\count@<10 0\fi\number\count@}}%
     \minute=\expandafter{\hack@}%
     \ifnum\count@<12
          \amorpm={am}
     \else\amorpm={pm}
          \ifnum\count@>12 \advance\count@ by-12 \fi
     \fi
     \standardhour=\expandafter{\number\count@}%
     \def\hack@19##1##2{\shortyear={##1##2}}%
          \expandafter\hack@\the\year}

\def\monthword#1{%
     \ifcase#1
          $\bullet$\err@badcountervalue{monthword}%
          \or January\or February\or March\or April\or May\or June%
          \or July\or August\or September\or October\or November\or December%
     \else$\bullet$\err@badcountervalue{monthword}%
     \fi}

\def\monthabbr#1{%
     \ifcase#1
          $\bullet$\err@badcountervalue{monthabbr}%
          \or Jan\or Feb\or Mar\or Apr\or May\or Jun%
          \or Jul\or Aug\or Sep\or Oct\or Nov\or Dec%
     \else$\bullet$\err@badcountervalue{monthabbr}%
     \fi}

\def\militarytime{\the\militaryhour:\the\minute}
\def\standardtime{\the\standardhour:\the\minute}

%************** Number styles *******************************

\def\@setnumstyle#1#2{\expandafter\global\expandafter\expandafter
     \expandafter\let\expandafter\expandafter
     \csname @\expandafter\blank\string#1style\endcsname
     \csname#2\endcsname}
\def\numstyle#1{\usename{@\expandafter\blank\string#1style}#1}
\def\ifblank#1\then{\useafter\ifx{@\expandafter\blank\string#1}\blank}

\def\blank#1{}

\def\Roman#1{\expandafter\uppercase\expandafter{\romannumeral#1}}
\def\alphabetic#1{%
     \ifcase#1
          $\bullet$\err@badcountervalue{alphabetic}%
          \or a\or b\or c\or d\or e\or f\or g\or h\or i\or j\or k\or l\or m%
          \or n\or o\or p\or q\or r\or s\or t\or u\or v\or w\or x\or y\or z%
     \else$\bullet$\err@badcountervalue{alphabetic}%
     \fi}
\def\Alphabetic#1{\expandafter\uppercase\expandafter{\alphabetic{#1}}}
\def\symbols#1{%
     \ifcase#1
          $\bullet$\err@badcountervalue{symbols}%
          \or*\or\dag\or\ddag\or\S\or$\|$%
          \or**\or\dag\dag\or\ddag\ddag\or\S\S\or$\|\|$%
     \else$\bullet$\err@badcountervalue{symbols}%
     \fi}

%************** String macros *******************************

\catcode`\^^?=13 \def^^?{\relax}

\def\trimleading#1\to#2{\edef#2{#1}%
     \expandafter\@trimleading\expandafter#2#2^^?^^?}
\def\@trimleading#1#2#3^^?{\ifx#2^^?\def#1{}\else\def#1{#2#3}\fi}

\def\trimtrailing#1\to#2{\edef#2{#1}%
     \expandafter\@trimtrailing\expandafter#2#2^^? ^^?\relax}
\def\@trimtrailing#1#2 ^^?#3{\ifx#3\relax\toks@={}%
     \else\def#1{#2}\toks@={\trimtrailing#1\to#1}\fi
     \the\toks@}

\def\trim#1\to#2{\trimleading#1\to#2\trimtrailing#2\to#2}

\catcode`\^^?=15

%************** List macros *********************************

\long\def\additemL#1\to#2{\toks@={\^^\{#1}}\toks@ii=\expandafter{#2}%
     \xdef#2{\the\toks@\the\toks@ii}}

\long\def\additemR#1\to#2{\toks@={\^^\{#1}}\toks@ii=\expandafter{#2}%
     \xdef#2{\the\toks@ii\the\toks@}}

\def\getitemL#1\to#2{\expandafter\@getitemL#1\hack@#1#2}
\def\@getitemL\^^\#1#2\hack@#3#4{\def#4{#1}\def#3{#2}}

%************************************************************
%*
%*             Font-related macros
%*
%************************************************************
\message{font macros,}

%************** Font set-up *********************************

\newdimen\rp@
\newcount\@@sizeindex \@@sizeindex=0
\newcount\@@factori
\newcount\@@factorii
\newcount\@@factoriii
\newcount\@@factoriv

\countdef\maxfam=18
\newfam\itfam
\newfam\bffam
\newfam\bfsfam
\newfam\bmitfam

\def\@mathfontinit{\count@=4
     \loop\textfont\count@=\nullfont
          \scriptfont\count@=\nullfont
          \scriptscriptfont\count@=\nullfont
          \ifnum\count@<\maxfam\advance\count@ by\@ne
     \repeat}

\def\@fontstyleinit{%
     \def\it{\err@fontnotavailable\it}%
     \def\bf{\err@fontnotavailable\bf}%
     \def\bfs{\err@bfstobf}%
     \def\bmit{\err@fontnotavailable\bmit}%
     \def\sc{\err@fontnotavailable\sc}%
     \def\sl{\err@sltoit}%
     \def\ss{\err@fontnotavailable\ss}%
     \def\tt{\err@fontnotavailable\tt}}

\def\@parameterinit#1{\rm\rp@=.1em \@getscaling{#1}%
     \let\^^\=\@doscaling\scalingskipslist
     \setbox\strutbox=\hbox{\vrule
          height.708\baselineskip depth.292\baselineskip width\z@}}

\def\@getfactor#1#2#3#4{\@@factori=#1 \@@factorii=#2
     \@@factoriii=#3 \@@factoriv=#4}

\def\@getscaling#1{\count@=#1 \advance\count@ by-\@@sizeindex\@@sizeindex=#1
     \ifnum\count@<0
          \let\@mulordiv=\divide
          \let\@divormul=\multiply
          \multiply\count@ by\m@ne
     \else\let\@mulordiv=\multiply
          \let\@divormul=\divide
     \fi
     \edef\@@scratcha{\ifcase\count@                {1}{1}{1}{1}\or
          {1}{7}{23}{3}\or     {2}{5}{3}{1}\or      {9}{89}{13}{1}\or
          {6}{25}{6}{1}\or     {8}{71}{14}{1}\or    {6}{25}{36}{5}\or
          {1}{7}{53}{4}\or     {12}{125}{108}{5}\or {3}{14}{53}{5}\or
          {6}{41}{17}{1}\or    {13}{31}{13}{2}\or   {9}{107}{71}{2}\or
          {11}{139}{124}{3}\or {1}{6}{43}{2}\or     {10}{107}{42}{1}\or
          {1}{5}{43}{2}\or     {5}{69}{65}{1}\or    {11}{97}{91}{2}\fi}%
     \expandafter\@getfactor\@@scratcha}

\def\@doscaling#1{\@mulordiv#1by\@@factori\@divormul#1by\@@factorii
     \@mulordiv#1by\@@factoriii\@divormul#1by\@@factoriv}

%************* Size-changing commands ***********************

\newskip\headskip
\newskip\footskip

\def\typesize=#1pt{\count@=#1 \advance\count@ by-10
     \ifcase\count@
          \@setsizex\or\err@badtypesize\or
          \@setsizexii\or\err@badtypesize\or
          \@setsizexiv
     \else\err@badtypesize
     \fi}

\def\@setsizex{\getixpt
     \def\subsubscriptfonts{\vpt}%
          \def\subsubscriptsize{\vpt\@parameterinit{-8}}%
     \def\subscriptfonts{\viipt}\def\subscriptsize{\viipt\@parameterinit{-4}}%
     \def\footnotefonts{\viiipt}\def\footnotesize{\viiipt\@parameterinit{-2}}%
     \def\smallfonts{\ixpt}\def\smallsize{\ixpt\@parameterinit{-1}}%
     \def\normalfonts{\xpt}\def\normalsize{\xpt\@parameterinit{0}}%
     \def\bigfonts{\xiipt}\def\bigsize{\xiipt\@parameterinit{2}}%
     \def\Bigfonts{\xivpt}\def\Bigsize{\xivpt\@parameterinit{4}}%
     \def\biggfonts{\xviipt}\def\biggsize{\xviipt\@parameterinit{6}}%
     \def\Biggfonts{\xxipt}\def\Biggsize{\xxipt\@parameterinit{8}}%
     \def\tinyfonts{\vpt}\def\tinysize{\vpt\@parameterinit{-8}}%
     \def\HUGEFONTS{\xxvpt}\def\HUGESIZE{\xxvpt\@parameterinit{10}}%
     \normalsize\fixedskipslist}

\def\@setsizexii{\getxipt
     \def\subsubscriptfonts{\vipt}%
          \def\subsubscriptsize{\vipt\@parameterinit{-6}}%
     \def\subscriptfonts{\viiipt}%
          \def\subscriptsize{\viiipt\@parameterinit{-2}}%
     \def\footnotefonts{\xpt}\def\footnotesize{\xpt\@parameterinit{0}}%
     \def\smallfonts{\xipt}\def\smallsize{\xipt\@parameterinit{1}}%
     \def\normalfonts{\xiipt}\def\normalsize{\xiipt\@parameterinit{2}}%
     \def\bigfonts{\xivpt}\def\bigsize{\xivpt\@parameterinit{4}}%
     \def\Bigfonts{\xviipt}\def\Bigsize{\xviipt\@parameterinit{6}}%
     \def\biggfonts{\xxipt}\def\biggsize{\xxipt\@parameterinit{8}}%
     \def\Biggfonts{\xxvpt}\def\Biggsize{\xxvpt\@parameterinit{10}}%
     \def\tinyfonts{\vpt}\def\tinysize{\vpt\@parameterinit{-8}}%
     \def\HUGEFONTS{\xxvpt}\def\HUGESIZE{\xxvpt\@parameterinit{10}}%
     \normalsize\fixedskipslist}

\def\@setsizexiv{\getxiiipt
     \def\subsubscriptfonts{\viipt}%
          \def\subsubscriptsize{\viipt\@parameterinit{-4}}%
     \def\subscriptfonts{\xpt}\def\subscriptsize{\xpt\@parameterinit{0}}%
     \def\footnotefonts{\xiipt}\def\footnotesize{\xiipt\@parameterinit{2}}%
     \def\smallfonts{\xiiipt}\def\smallsize{\xiiipt\@parameterinit{3}}%
     \def\normalfonts{\xivpt}\def\normalsize{\xivpt\@parameterinit{4}}%
     \def\bigfonts{\xviipt}\def\bigsize{\xviipt\@parameterinit{6}}%
     \def\Bigfonts{\xxipt}\def\Bigsize{\xxipt\@parameterinit{8}}%
     \def\biggfonts{\xxvpt}\def\biggsize{\xxvpt\@parameterinit{10}}%
     \def\Biggfonts{\err@sizetoolarge\Biggfonts\HUGEFONTS}%
          \def\Biggsize{\err@sizetoolarge\Biggsize\HUGESIZE}%
     \def\tinyfonts{\vpt}\def\tinysize{\vpt\@parameterinit{-8}}%
     \def\HUGEFONTS{\xxvpt}\def\HUGESIZE{\xxvpt\@parameterinit{10}}%
     \normalsize\fixedskipslist}

\def\subsubscriptfonts{\vpt} \def\subsubscriptsize{\vpt\@parameterinit{-8}}
\def\subscriptfonts{\viipt}  \def\subscriptsize{\viipt\@parameterinit{-4}}
\def\footnotefonts{\viiipt}  \def\footnotesize{\viiipt\@parameterinit{-2}}
\def\smallfonts{\err@sizenotavailable\smallfonts}
                             \def\smallsize{\ixpt\@parameterinit{-1}}
\def\normalfonts{\xpt}       \def\normalsize{\xpt\@parameterinit{0}}
\def\bigfonts{\xiipt}        \def\bigsize{\xiipt\@parameterinit{2}}
\def\Bigfonts{\xivpt}        \def\Bigsize{\xivpt\@parameterinit{4}}
\def\biggfonts{\xviipt}      \def\biggsize{\xviipt\@parameterinit{6}}
\def\Biggfonts{\xxipt}       \def\Biggsize{\xxipt\@parameterinit{8}}
\def\tinyfonts{\vpt}         \def\tinysize{\vpt\@parameterinit{-8}}
\def\HUGEFONTS{\xxvpt}       \def\HUGESIZE{\xxvpt\@parameterinit{10}}

%************************************************************
%*
%*             Document layout
%*
%************************************************************
\message{document layout,}

%************** Page format *********************************

\newtoks\everyoutput \everyoutput={}
\newdimen\depthofpage
\newcount\pagenum \pagenum=0

\newdimen\oddtopmargin  \newdimen\eventopmargin
\newdimen\oddleftmargin \newdimen\evenleftmargin
\newtoks\oddhead        \newtoks\evenhead
\newtoks\oddfoot        \newtoks\evenfoot

\def\topmargin{\afterassignment\@seteventop\oddtopmargin}
\def\leftmargin{\afterassignment\@setevenleft\oddleftmargin}
\def\head{\afterassignment\@setevenhead\oddhead}
\def\foot{\afterassignment\@setevenfoot\oddfoot}

\def\@seteventop{\eventopmargin=\oddtopmargin}
\def\@setevenleft{\evenleftmargin=\oddleftmargin}
\def\@setevenhead{\evenhead=\oddhead}
\def\@setevenfoot{\evenfoot=\oddfoot}

\def\pagenumstyle#1{\@setnumstyle\pagenum{#1}}

\newif\ifdraft
\def\draft{\drafttrue\leftmargin=.5in \overfullrule=5pt }

\def\outputstyle#1{\global\expandafter\let\expandafter
          \@outputstyle\csname#1output\endcsname
     \usename{#1setup}}

\output={\@outputstyle}

\def\normaloutput{\the\everyoutput
     \global\advance\pagenum by\@ne
     \ifodd\pagenum
          \voffset=\oddtopmargin \hoffset=\oddleftmargin
     \else\voffset=\eventopmargin \hoffset=\evenleftmargin
     \fi
     \advance\voffset by-1in  \advance\hoffset by-1in
     \count0=\pagenum
     \expandafter\shipout\pagebox
     \ifnum\outputpenalty>-\@MM\else\dosupereject\fi}

\newdimen\fullhsize
\newbox\leftpage
\newcount\leftpagenum
\newcount\outputpagenum \outputpagenum=0
\let\leftorright=L

\def\twoupoutput{\the\everyoutput
     \global\advance\pagenum by\@ne
     \if L\leftorright
          \global\setbox\leftpage=\leftline{\pagebox}%
          \global\leftpagenum=\pagenum
          \global\let\leftorright=R%
     \else\global\advance\outputpagenum by\@ne
          \ifodd\outputpagenum
               \voffset=\oddtopmargin \hoffset=\oddleftmargin
          \else\voffset=\eventopmargin \hoffset=\evenleftmargin
          \fi
          \advance\voffset by-1in  \advance\hoffset by-1in
          \count0=\leftpagenum \count1=\pagenum
          \shipout\vbox{\hbox to\fullhsize
               {\box\leftpage\hfil\leftline{\pagebox}}}%
          \global\let\leftorright=L%
     \fi
     \ifnum\outputpenalty>-\@MM
     \else\dosupereject
          \if R\leftorright
               \globaldefs=\@ne\head={\hfil}\foot={\hfil}\globaldefs=\z@
               \null\newpage
          \fi
     \fi}

\def\pagebox{\vbox{\makeheadline\pagebody\makefootline}}

\def\makeheadline{%
     \vbox to\z@{\baselinestretch=\@m
          \vskip\topskip\vskip-.708\baselineskip\vskip-\headskip
          \line{\vbox to\ht\strutbox{}%
               \ifodd\pagenum\the\oddhead\else\the\evenhead\fi}%
          \vss}%
     \nointerlineskip}

\def\pagebody{\vbox to\vsize{%
     \boxmaxdepth\maxdepth
     \ifvoid\topins\else\unvbox\topins\fi
     \depthofpage=\dp255
     \unvbox255
     \ifraggedbottom\kern-\depthofpage\vfil\fi
     \ifvoid\footins
     \else\vskip\skip\footins
          \footnoterule
          \unvbox\footins
          \vskip-\footnoteskip
     \fi}}

\def\makefootline{\baselineskip=\footskip
     \line{\ifodd\pagenum\the\oddfoot\else\the\evenfoot\fi}}

%************** Sectioning commands *************************

\newskip\abovechapterskip
\newskip\belowchapterskip
\newskip\abovesectionskip
\newskip\belowsectionskip
\newskip\abovesubsectionskip
\newskip\belowsubsectionskip

\def\chapterstyle#1{\global\expandafter\let\expandafter\@chapterstyle
     \csname#1text\endcsname}
\def\sectionstyle#1{\global\expandafter\let\expandafter\@sectionstyle
     \csname#1text\endcsname}
\def\subsectionstyle#1{\global\expandafter\let\expandafter\@subsectionstyle
     \csname#1text\endcsname}

\def\chapter#1{%
     \ifdim\lastskip=17sp \else\chapterbreak\vskip\abovechapterskip\fi
     \@chapterstyle{\ifblank\chapternumstyle\then
          \else\newchapternum=\next\chapternumformat\ \fi#1}%
     \nobreak\vskip\belowchapterskip\vskip17sp }

\def\section#1{%
     \ifdim\lastskip=17sp \else\sectionbreak\vskip\abovesectionskip\fi
     \@sectionstyle{\ifblank\sectionnumstyle\then
          \else\newsectionnum=\next\sectionnumformat\ \fi#1}%
     \nobreak\vskip\belowsectionskip\vskip17sp }

\def\subsection#1{%
     \ifdim\lastskip=17sp \else\subsectionbreak\vskip\abovesubsectionskip\fi
     \@subsectionstyle{\ifblank\subsectionnumstyle\then
          \else\newsubsectionnum=\next\subsectionnumformat\ \fi#1}%
     \nobreak\vskip\belowsubsectionskip\vskip17sp }

%************** Text formatting commands ********************

\let\TeXunderline=\underline
\let\TeXoverline=\overline
\def\underline#1{\relax\ifmmode\TeXunderline{#1}\else
     $\TeXunderline{\hbox{#1}}$\fi}
\def\overline#1{\relax\ifmmode\TeXoverline{#1}\else
     $\TeXoverline{\hbox{#1}}$\fi}

\def\baselinestretch{\afterassignment\@baselinestretch\count@}
\def\@baselinestretch{\baselineskip=\normalbaselineskip
     \divide\baselineskip by\@m\baselineskip=\count@\baselineskip
     \setbox\strutbox=\hbox{\vrule
          height.708\baselineskip depth.292\baselineskip width\z@}%
     \bigskipamount=\the\baselineskip
          plus.25\baselineskip minus.25\baselineskip
     \medskipamount=.5\baselineskip
          plus.125\baselineskip minus.125\baselineskip
     \smallskipamount=.25\baselineskip
          plus.0625\baselineskip minus.0625\baselineskip}

\def\\{\ifhmode\ifnum\lastpenalty=-\@M\else\hfil\penalty-\@M\fi\fi
     \ignorespaces}
\def\newpage{\vfil\break}

\def\lefttext#1{\par{\@text\leftskip=\z@\rightskip=\centering
     \noindent#1\par}}
\def\righttext#1{\par{\@text\leftskip=\centering\rightskip=\z@
     \noindent#1\par}}
\def\centertext#1{\par{\@text\leftskip=\centering\rightskip=\centering
     \noindent#1\par}}
\def\@text{\parindent=\z@ \parfillskip=\z@ \everypar={}%
     \spaceskip=.3333em \xspaceskip=.5em
     \def\\{\ifhmode\ifnum\lastpenalty=-\@M\else\penalty-\@M\fi\fi
          \ignorespaces}}

\def\beginleft{\par\@text\leftskip=\z@ \rightskip=\centering}
     
\def\beginright{\par\@text\leftskip=\centering\rightskip=\z@ }
     
\def\begincenter{\par\@text\leftskip=\centering\rightskip=\centering}

\def\beginnarrow{\defaultoption[\parindent]\@beginnarrow}
\def\@beginnarrow[#1]{\par\advance\leftskip by#1\advance\rightskip by#1}

\begingroup
\catcode`\[=1 \catcode`\{=11 \gdef\beginignore[\endgroup\bgroup
     \catcode`\e=0 \catcode`\\=12 \catcode`\{=11 \catcode`\f=12 \let\or=\relax
     \let\nd{ignor=\fi \let\}=\egroup
     \iffalse}
\endgroup

\long\def\marginnote#1{\leavevmode
     \edef\@marginsf{\spacefactor=\the\spacefactor\relax}%
     \ifdraft\strut\vadjust{%
          \hbox to\z@{\hskip\hsize\hskip.1in
               \vbox to\z@{\vskip-\dp\strutbox
                    \marginnoteformat
                    \vskip-\ht\strutbox
                    \noindent\strut#1\par
                    \vss}%
               \hss}}%
     \fi
     \@marginsf}

%************** The \bye command ****************************

\newtoks\everybye \everybye={\par\vfil}
\outer\def\bye{\the\everybye
     \footnotecheck
     \prelabelcheck
     \streamcheck
     \supereject
     \TeXend}

%************************************************************
%*
%*             Footnotes
%*
%************************************************************
\message{footnotes,}

\newcount\footnotenum \footnotenum=0
\newskip\footnoteskip
\let\@footnotelist=\empty

\def\footnotenumstyle#1{\@setnumstyle\footnotenum{#1}%
     \useafter\ifx{@footnotenumstyle}\symbols
          \global\let\@footup=\empty
     \else\global\let\@footup=\markup
     \fi}

\def\footnote{\footnotecheck\defaultoption[]\@footnote}
\def\@footnote[#1]{\@footnotemark[#1]\@footnotetext}

\def\footnotemark{\defaultoption[]\@footnotemark}
\def\@footnotemark[#1]{\let\@footsf=\empty
     \ifhmode\edef\@footsf{\spacefactor=\the\spacefactor\relax}\/\fi
     \ifnoarg#1\then
          \global\advance\footnotenum by\@ne
          \@footup{\footnotenumformat}%
          \edef\@@foota{\footnotenum=\the\footnotenum\relax}%
          \expandafter\additemR\expandafter\@footup\expandafter
               {\@@foota\footnotenumformat}\to\@footnotelist
          \global\let\@footnotelist=\@footnotelist
     \else\markup{#1}%
          \additemR\markup{#1}\to\@footnotelist
          \global\let\@footnotelist=\@footnotelist
     \fi
     \@footsf}

\def\footnotetext{%
     \ifx\@footnotelist\empty\err@extrafootnotetext\else\@footnotetext\fi}
\def\@footnotetext{%
     \getitemL\@footnotelist\to\@@foota
     \global\let\@footnotelist=\@footnotelist
     \insert\footins\bgroup
     \footnoteformat
     \splittopskip=\ht\strutbox\splitmaxdepth=\dp\strutbox
     \interlinepenalty=\interfootnotelinepenalty\floatingpenalty=\@MM
     \noindent\llap{\@@foota}\strut
     \bgroup\aftergroup\@footnoteend
     \let\@@scratcha=}
\def\@footnoteend{\strut\par\vskip\footnoteskip\egroup}

\def\footnoterule{\normalfonts
     \kern-.3em \hrule width2in height.04em \kern .26em }

\def\footnotecheck{%
     \ifx\@footnotelist\empty
     \else\err@extrafootnotemark
          \global\let\@footnotelist=\empty
     \fi}

%************************************************************
%*
%*             Labelling macros
%*
%************************************************************
\message{labels,}

\let\@@labeldef=\xdef
\newif\if@labelfile
\newwrite\@labelfile
\let\@prelabellist=\empty

\def\label#1#2{\trim#1\to\@@labarg\edef\@@labtext{#2}%
     \edef\@@labname{lab@\@@labarg}%
     \useafter\ifundefined\@@labname\then\else\@yeslab\fi
     \useafter\@@labeldef\@@labname{#2}%
     \ifstreaming
          \expandafter\toks@\expandafter\expandafter\expandafter
               {\csname\@@labname\endcsname}%
          \immediate\write\streamout{\noexpand\label{\@@labarg}{\the\toks@}}%
     \fi}
\def\@yeslab{%
     \useafter\ifundefined{if\@@labname}\then
          \err@labelredef\@@labarg
     \else\useif{if\@@labname}\then
               \err@labelredef\@@labarg
          \else\global\usename{\@@labname true}%
               \useafter\ifundefined{pre\@@labname}\then
               \else\useafter\ifx{pre\@@labname}\@@labtext
                    \else\err@badlabelmatch\@@labarg
                    \fi
               \fi
               \if@labelfile
               \else\global\@labelfiletrue
                    \immediate\write\sixt@@n{--> Creating file \jobname.lab}%
                    \immediate\openout\@labelfile=\jobname.lab
               \fi
               \immediate\write\@labelfile
                    {\noexpand\prelabel{\@@labarg}{\@@labtext}}%
          \fi
     \fi}

\def\putlab#1{\trim#1\to\@@labarg\edef\@@labname{lab@\@@labarg}%
     \useafter\ifundefined\@@labname\then\@nolab\else\usename\@@labname\fi}
\def\@nolab{%
     \useafter\ifundefined{pre\@@labname}\then
          \undefinedlabelformat
          \err@needlabel\@@labarg
          \useafter\xdef\@@labname{\undefinedlabelformat}%
     \else\usename{pre\@@labname}%
          \useafter\xdef\@@labname{\usename{pre\@@labname}}%
     \fi
     \useafter\newif{if\@@labname}%
     \expandafter\additemR\@@labarg\to\@prelabellist}

\def\prelabel#1{\useafter\gdef{prelab@#1}}

\def\ifundefinedlabel#1\then{%
     \expandafter\ifx\csname lab@#1\endcsname\relax}
\def\useiflab#1\then{\csname iflab@#1\endcsname}

\def\prelabelcheck{{%
     \def\^^\##1{\useiflab{##1}\then\else\err@undefinedlabel{##1}\fi}%
     \@prelabellist}}

%************************************************************
%*
%*             Equation numbering
%*
%************************************************************
\message{equation numbering,}

\newcount\chapternum
\newcount\sectionnum
\newcount\subsectionnum
\newcount\equationnum
\newcount\subequationnum
\newcount\figurenum
\newcount\subfigurenum
\newcount\tablenum
\newcount\subtablenum

\newif\if@subeqncount
\newif\if@subfigcount
\newif\if@subtblcount

\def\newchapternum{\newsectionnum=\z@\@resetnum\chapternum}
\def\newsectionnum{\newsubsectionnum=\z@\@resetnum\sectionnum}
\def\newsubsectionnum{\newequationnum=\z@\newfigurenum=\z@\newtablenum=\z@
     \@resetnum\subsectionnum}
\def\newequationnum{\newsubequationnum=\z@\@resetnum\equationnum}
\def\newsubequationnum{\@resetnum\subequationnum}
\def\newfigurenum{\newsubfigurenum=\z@\@resetnum\figurenum}
\def\newsubfigurenum{\@resetnum\subfigurenum}
\def\newtablenum{\newsubtablenum=\z@\@resetnum\tablenum}
\def\newsubtablenum{\@resetnum\subtablenum}

\def\@resetnum#1{\global\advance#1by1 \edef\next{\the#1\relax}\global#1}

\newchapternum=0

\def\chapternumstyle#1{\@setnumstyle\chapternum{#1}}
\def\sectionnumstyle#1{\@setnumstyle\sectionnum{#1}}
\def\subsectionnumstyle#1{\@setnumstyle\subsectionnum{#1}}
\def\equationnumstyle#1{\@setnumstyle\equationnum{#1}}
\def\subequationnumstyle#1{\@setnumstyle\subequationnum{#1}%
     \ifblank\subequationnumstyle\then\global\@subeqncountfalse\fi
     \ignorespaces}
\def\figurenumstyle#1{\@setnumstyle\figurenum{#1}}
\def\subfigurenumstyle#1{\@setnumstyle\subfigurenum{#1}%
     \ifblank\subfigurenumstyle\then\global\@subfigcountfalse\fi
     \ignorespaces}
\def\tablenumstyle#1{\@setnumstyle\tablenum{#1}}
\def\subtablenumstyle#1{\@setnumstyle\subtablenum{#1}%
     \ifblank\subtablenumstyle\then\global\@subtblcountfalse\fi
     \ignorespaces}

\def\eqnlabel#1{%
     \if@subeqncount
          \newsubequationnum=\next
     \else\newequationnum=\next
          \ifblank\subequationnumstyle\then
          \else\global\@subeqncounttrue
               \newsubequationnum=\@ne
          \fi
     \fi
     \label{#1}{\puteqnformat}(\puteqn{#1})%
     \ifdraft\rlap{\hskip.1in{\tt#1}}\fi}

\let\puteqn=\putlab

\def\equation#1#2{\useafter\gdef{eqn@#1}{#2\eqno\eqnlabel{#1}}}
\def\Equation#1{\useafter\gdef{eqn@#1}}

\def\putequation#1{\useafter\ifundefined{eqn@#1}\then
     \err@undefinedeqn{#1}\else\usename{eqn@#1}\fi}

\def\eqnseriesstyle#1{\gdef\@eqnseriesstyle{#1}}
\def\begineqnseries{\subequationnumstyle{\@eqnseriesstyle}%
     \defaultoption[]\@begineqnseries}
\def\@begineqnseries[#1]{\edef\@@eqnname{#1}}
\def\endeqnseries{\subequationnumstyle{blank}%
     \expandafter\ifnoarg\@@eqnname\then
     \else\label\@@eqnname{\puteqnformat}%
     \fi
     \aftergroup\ignorespaces}

\def\figlabel#1{%
     \if@subfigcount
          \newsubfigurenum=\next
     \else\newfigurenum=\next
          \ifblank\subfigurenumstyle\then
          \else\global\@subfigcounttrue
               \newsubfigurenum=\@ne
          \fi
     \fi
     \label{#1}{\putfigformat}\putfig{#1}%
     {\def\marginnoteformat{\tt}\marginnote{#1}}}

\let\putfig=\putlab

\def\figseriesstyle#1{\gdef\@figseriesstyle{#1}}
\def\beginfigseries{\subfigurenumstyle{\@figseriesstyle}%
     \defaultoption[]\@beginfigseries}
\def\@beginfigseries[#1]{\edef\@@figname{#1}}
\def\endfigseries{\subfigurenumstyle{blank}%
     \expandafter\ifnoarg\@@figname\then
     \else\label\@@figname{\putfigformat}%
     \fi
     \aftergroup\ignorespaces}

\def\tbllabel#1{%
     \if@subtblcount
          \newsubtablenum=\next
     \else\newtablenum=\next
          \ifblank\subtablenumstyle\then
          \else\global\@subtblcounttrue
               \newsubtablenum=\@ne
          \fi
     \fi
     \label{#1}{\puttblformat}\puttbl{#1}%
     {\def\marginnoteformat{\tt}\marginnote{#1}}}

\let\puttbl=\putlab

\def\tblseriesstyle#1{\gdef\@tblseriesstyle{#1}}
\def\begintblseries{\subtablenumstyle{\@tblseriesstyle}%
     \defaultoption[]\@begintblseries}
\def\@begintblseries[#1]{\edef\@@tblname{#1}}
\def\endtblseries{\subtablenumstyle{blank}%
     \expandafter\ifnoarg\@@tblname\then
     \else\label\@@tblname{\puttblformat}%
     \fi
     \aftergroup\ignorespaces}

%************************************************************
%*
%*             Reference numbering
%*
%************************************************************
\message{reference numbering,}

\newcount\referencenum \referencenum=0
\newcount\@@prerefcount \@@prerefcount=0
\newcount\@@thisref
\newcount\@@lastref
\newcount\@@loopref
\newcount\@@refseq
\newdimen\refnumindent
\let\@undefreflist=\empty

\def\referencenumstyle#1{\@setnumstyle\referencenum{#1}}

\def\referencestyle#1{\usename{@ref#1}}

\def\@refsequential{%
     \gdef\@refpredef##1{\global\advance\referencenum by\@ne
          \let\^^\=0\label{##1}{\^^\{\the\referencenum}}%
          \useafter\gdef{ref@\the\referencenum}{{##1}{\undefinedlabelformat}}}%
     \gdef\@reference##1##2{%
          \ifundefinedlabel##1\then
          \else\def\^^\####1{\global\@@thisref=####1\relax}\putlab{##1}%
               \useafter\gdef{ref@\the\@@thisref}{{##1}{##2}}%
          \fi}%
     \gdef\endputreferences{%
          \loop\ifnum\@@loopref<\referencenum
                    \advance\@@loopref by\@ne
                    \expandafter\expandafter\expandafter\@printreference
                         \csname ref@\the\@@loopref\endcsname
          \repeat
          \par}}

\def\@refpreordered{%
     \gdef\@refpredef##1{\global\advance\referencenum by\@ne
          \additemR##1\to\@undefreflist}%
     \gdef\@reference##1##2{%
          \ifundefinedlabel##1\then
          \else\global\advance\@@loopref by\@ne
               {\let\^^\=0\label{##1}{\^^\{\the\@@loopref}}}%
               \@printreference{##1}{##2}%
          \fi}
     \gdef\endputreferences{%
          \def\^^\####1{\useiflab{####1}\then
               \else\reference{####1}{\undefinedlabelformat}\fi}%
          \@undefreflist
          \par}}

\def\beginprereferences{\par
     \def\reference##1##2{\global\advance\referencenum by1\@ne
          \let\^^\=0\label{##1}{\^^\{\the\referencenum}}%
          \useafter\gdef{ref@\the\referencenum}{{##1}{##2}}}}
\def\endprereferences{\global\@@prerefcount=\the\referencenum\par}

\def\beginputreferences{\par
     \refnumindent=\z@\@@loopref=\z@
     \loop\ifnum\@@loopref<\referencenum
               \advance\@@loopref by\@ne
               \setbox\z@=\hbox{\referencenum=\@@loopref
                    \referencenumformat\enskip}%
               \ifdim\wd\z@>\refnumindent\refnumindent=\wd\z@\fi
     \repeat
     \putreferenceformat
     \@@loopref=\z@
     \loop\ifnum\@@loopref<\@@prerefcount
               \advance\@@loopref by\@ne
               \expandafter\expandafter\expandafter\@printreference
                    \csname ref@\the\@@loopref\endcsname
     \repeat
     \let\reference=\@reference}

\def\@printreference#1#2{\ifx#2\undefinedlabelformat\err@undefinedref{#1}\fi
     \noindent\ifdraft\rlap{\hskip\hsize\hskip.1in \tt#1}\fi
     \llap{\referencenum=\@@loopref\referencenumformat\enskip}#2\par}

\def\reference#1#2{{\par\refnumindent=\z@\putreferenceformat\noindent#2\par}}

\def\putref#1{\trim#1\to\@@refarg
     \expandafter\ifnoarg\@@refarg\then
          \toks@={\relax}%
     \else\@@lastref=-\@m\def\@@refsep{}\def\@more{\@nextref}%
          \toks@={\@nextref#1,,}%
     \fi\the\toks@}
\def\@nextref#1,{\trim#1\to\@@refarg
     \expandafter\ifnoarg\@@refarg\then
          \let\@more=\relax
     \else\ifundefinedlabel\@@refarg\then
               \expandafter\@refpredef\expandafter{\@@refarg}%
          \fi
          \def\^^\##1{\global\@@thisref=##1\relax}%
          \global\@@thisref=\m@ne
          \setbox\z@=\hbox{\putlab\@@refarg}%
     \fi
     \advance\@@lastref by\@ne
     \ifnum\@@lastref=\@@thisref\advance\@@refseq by\@ne\else\@@refseq=\@ne\fi
     \ifnum\@@lastref<\z@
     \else\ifnum\@@refseq<\thr@@
               \@@refsep\def\@@refsep{,}%
               \ifnum\@@lastref>\z@
                    \advance\@@lastref by\m@ne
                    {\referencenum=\@@lastref\putrefformat}%
               \else\undefinedlabelformat
               \fi
          \else\def\@@refsep{--}%
          \fi
     \fi
     \@@lastref=\@@thisref
     \@more}

%************************************************************
%*
%*             Job streaming
%*
%************************************************************
\message{streaming,}

\newif\ifstreaming

\def\streamto{\defaultoption[\jobname]\@streamto}
\def\@streamto[#1]{\global\streamingtrue
     \immediate\write\sixt@@n{--> Streaming to #1.str}%
     \newwrite\streamout\immediate\openout\streamout=#1.str }

\def\streamfrom{\defaultoption[\jobname]\@streamfrom}
\def\@streamfrom[#1]{\newread\streamin\openin\streamin=#1.str
     \ifeof\streamin
          \expandafter\err@nostream\expandafter{#1.str}%
     \else\immediate\write\sixt@@n{--> Streaming from #1.str}%
          \let\@@labeldef=\gdef
          \ifstreaming
               \edef\@elc{\endlinechar=\the\endlinechar}%
               \endlinechar=\m@ne
               \loop\read\streamin to\@@scratcha
                    \ifeof\streamin
                         \streamingfalse
                    \else\toks@=\expandafter{\@@scratcha}%
                         \immediate\write\streamout{\the\toks@}%
                    \fi
                    \ifstreaming
               \repeat
               \@elc
               \input #1.str
               \streamingtrue
          \else\input #1.str
          \fi
          \let\@@labeldef=\xdef
     \fi}

\def\streamcheck{\ifstreaming
     \immediate\write\streamout{\pagenum=\the\pagenum}%
     \immediate\write\streamout{\footnotenum=\the\footnotenum}%
     \immediate\write\streamout{\referencenum=\the\referencenum}%
     \immediate\write\streamout{\chapternum=\the\chapternum}%
     \immediate\write\streamout{\sectionnum=\the\sectionnum}%
     \immediate\write\streamout{\subsectionnum=\the\subsectionnum}%
     \immediate\write\streamout{\equationnum=\the\equationnum}%
     \immediate\write\streamout{\subequationnum=\the\subequationnum}%
     \immediate\write\streamout{\figurenum=\the\figurenum}%
     \immediate\write\streamout{\subfigurenum=\the\subfigurenum}%
     \immediate\write\streamout{\tablenum=\the\tablenum}%
     \immediate\write\streamout{\subtablenum=\the\subtablenum}%
     \immediate\closeout\streamout
     \fi}

%************************************************************
%*
%*             Error messages
%*
%************************************************************

\def\err@badtypesize{%
     \errhelp={The limited availability of certain fonts requires^^J%
          that the base type size be 10pt, 12pt, or 14pt.^^J}%
     \errmessage{--> Illegal base type size}}

\def\err@badsizechange{\immediate\write\sixt@@n
     {--> Size change not allowed in math mode, ignored}}

\def\err@sizetoolarge#1{\immediate\write\sixt@@n
     {--> \noexpand#1 too big, substituting HUGE}}

\def\err@sizenotavailable#1{\immediate\write\sixt@@n
     {--> Size not available, \noexpand#1 ignored}}

\def\err@fontnotavailable#1{\immediate\write\sixt@@n
     {--> Font not available, \noexpand#1 ignored}}

\def\err@sltoit{\immediate\write\sixt@@n
     {--> Style \noexpand\sl not available, substituting \noexpand\it}%
     \it}

\def\err@bfstobf{\immediate\write\sixt@@n
     {--> Style \noexpand\bfs not available, substituting \noexpand\bf}%
     \bf}

\def\err@badgroup#1#2{%
     \errhelp={The block you have just tried to close was not the one^^J%
          most recently opened.^^J}%
     \errmessage{--> \noexpand\end{#1} doesn't match \noexpand\begin{#2}}}

\def\err@badcountervalue#1{\immediate\write\sixt@@n
     {--> Counter (#1) out of bounds}}

\def\err@extrafootnotemark{\immediate\write\sixt@@n
     {--> \noexpand\footnotemark command
          has no corresponding \noexpand\footnotetext}}

\def\err@extrafootnotetext{%
     \errhelp{You have given a \noexpand\footnotetext command without first
          specifying^^Ja \noexpand\footnotemark.^^J}%
     \errmessage{--> \noexpand\footnotetext command has no corresponding
          \noexpand\footnotemark}}

\def\err@labelredef#1{\immediate\write\sixt@@n
     {--> Label "#1" redefined}}

\def\err@badlabelmatch#1{\immediate\write\sixt@@n
     {--> Definition of label "#1" doesn't match value in \jobname.lab}}

\def\err@needlabel#1{\immediate\write\sixt@@n
     {--> Label "#1" cited before its definition}}

\def\err@undefinedlabel#1{\immediate\write\sixt@@n
     {--> Label "#1" cited but never defined}}

\def\err@undefinedeqn#1{\immediate\write\sixt@@n
     {--> Equation "#1" not defined}}

\def\err@undefinedref#1{\immediate\write\sixt@@n
     {--> Reference "#1" not defined}}

\def\err@nostream#1{%
     \errhelp={You have tried to input a stream file that doesn't exist.^^J}%
     \errmessage{--> Stream file #1 not found}}

%************************************************************
%*
%*             Initialization
%*
%************************************************************
\message{jyTeX initialization}

\everyjob{\immediate\write16{--> jyTeX version \fmtversion}%
     \edef\@@jobname{\jobname}%
%     \openin0=\inputpath jysupp
%     \ifeof0
%     \else\closein0
%          \immediate\write16{--> Additional macros loaded from jysupp.tex}%
%          \jyinput jysupp
%     \fi
%     \openin0=\inputpath jylocal
%     \ifeof0
%     \else\closein0
%          \immediate\write16{--> Additional macros loaded from jylocal.tex}%
%          \jyinput jylocal
%     \fi
     \edef\jobname{\@@jobname}%
     \settime
     \openin0=\jobname.lab
     \ifeof0
     \else\closein0
          \immediate\write16{--> Getting labels from file \jobname.lab}%
          \input\jobname.lab
     \fi}

%************** Spacing *************************************

\def\fixedskipslist{%
     \^^\{\topskip}%
     \^^\{\splittopskip}%
     \^^\{\maxdepth}%
     \^^\{\skip\topins}%
     \^^\{\skip\footins}%
     \^^\{\headskip}%
     \^^\{\footskip}}

\def\scalingskipslist{%
     \^^\{\p@renwd}%
     \^^\{\delimitershortfall}%
     \^^\{\nulldelimiterspace}%
     \^^\{\scriptspace}%
     \^^\{\jot}%
     \^^\{\normalbaselineskip}%
     \^^\{\normallineskip}%
     \^^\{\normallineskiplimit}%
     \^^\{\baselineskip}%
     \^^\{\lineskip}%
     \^^\{\lineskiplimit}%
     \^^\{\bigskipamount}%
     \^^\{\medskipamount}%
     \^^\{\smallskipamount}%
     \^^\{\parskip}%
     \^^\{\parindent}%
     \^^\{\abovedisplayskip}%
     \^^\{\belowdisplayskip}%
     \^^\{\abovedisplayshortskip}%
     \^^\{\belowdisplayshortskip}%
     \^^\{\abovechapterskip}%
     \^^\{\belowchapterskip}%
     \^^\{\abovesectionskip}%
     \^^\{\belowsectionskip}%
     \^^\{\abovesubsectionskip}%
     \^^\{\belowsubsectionskip}}

%************** Document layout *****************************

\def\twoupsetup{%                                % setup for twoup style
     \topmargin=.75in
     \leftmargin=.5in
     \vsize=6.9in
     \hsize=4.75in
     \fullhsize=10in
     \let\draft=\relax}

\outputstyle{normal}                             % page style

\def\marginnoteformat{\subscriptsize             % paragraphing of margin notes
     \hsize=1in \baselinestretch=1000 \everypar={}%
     \tolerance=5000 \hbadness=5000 \parskip=0pt \parindent=0pt
     \leftskip=0pt \rightskip=0pt \raggedright}

\head={\ifdraft\normalfonts\it\hfil DRAFT\hfil   % format of headline
     \llap{\number\day\ \monthword\month\ \militarytime}\else\hfil\fi}
\foot={\hfil\normalfonts\numstyle\pagenum\hfil}  % format of footline

\normalbaselineskip=12pt                         % usual \baselineskip
\normallineskip=0pt                              % usual \lineskip
\normallineskiplimit=0pt                         % usual \lineskiplimit
\normalbaselines                                 % set \baselineskip

\topskip=.85\baselineskip \splittopskip=\topskip \headskip=2\baselineskip
\footskip=\headskip

\pagenumstyle{arabic}                            % counter style

\parskip=0pt                                     % no skip between paragraphs
\parindent=20pt                                  % usual \parindent

\baselinestretch=1000                            % set \big-, \med-, \smallskip

%************** Sectioning **********************************

\chapterstyle{left}                              % position of heading
\chapternumstyle{blank}                          % counter style
\def\chapterbreak{\newpage}                      % break before heading
\abovechapterskip=0pt                            % space before heading
\belowchapterskip=1.5\baselineskip               % space after heading
     plus.38\baselineskip minus.38\baselineskip
\def\chapternumformat{\numstyle\chapternum.}     % format of heading counter

\sectionstyle{left}                              % position of heading
\sectionnumstyle{blank}                          % counter style
\def\sectionbreak{\vskip0pt plus4\baselineskip\penalty-100
     \vskip0pt plus-4\baselineskip}              % break before heading
\abovesectionskip=1.5\baselineskip               % space before heading
     plus.38\baselineskip minus.38\baselineskip
\belowsectionskip=\the\baselineskip              % space after heading
     plus.25\baselineskip minus.25\baselineskip
\def\sectionnumformat{%                          % format of heading counter
     \ifblank\chapternumstyle\then\else\numstyle\chapternum.\fi
     \numstyle\sectionnum.}

\subsectionstyle{left}                           % position of heading
\subsectionnumstyle{blank}                       % counter style
\def\subsectionbreak{\vskip0pt plus4\baselineskip\penalty-100
     \vskip0pt plus-4\baselineskip}              % break before heading
\abovesubsectionskip=\the\baselineskip           % space before heading
     plus.25\baselineskip minus.25\baselineskip
\belowsubsectionskip=.75\baselineskip            % space after heading
     plus.19\baselineskip minus.19\baselineskip
\def\subsectionnumformat{%                       % format of heading counter
     \ifblank\chapternumstyle\then\else\numstyle\chapternum.\fi
     \ifblank\sectionnumstyle\then\else\numstyle\sectionnum.\fi
     \numstyle\subsectionnum.}

%************** Footnotes ***********************************

\footnotenumstyle{symbol}                       % counter style
\footnoteskip=0pt                                % jyTeX spacing parameter
\def\footnotenumformat{\numstyle\footnotenum}    % \footnotemark format
\def\footnoteformat{\footnotesize                % paragraphing of text
     \everypar={}\parskip=0pt \parfillskip=0pt plus1fil
     \leftskip=1em \rightskip=0pt
     \spaceskip=0pt \xspaceskip=0pt
     \def\\{\ifhmode\ifnum\lastpenalty=-10000
          \else\hfil\penalty-10000 \fi\fi\ignorespaces}}

%************** Labels **************************************

\def\undefinedlabelformat{$\bullet$}             % mark for undefined label

%************** Equation numbering **************************

\equationnumstyle{arabic}                        % counter style
\subequationnumstyle{blank}                      % counter style
\figurenumstyle{arabic}                          % counter style
\subfigurenumstyle{blank}                        % counter style
\tablenumstyle{arabic}                           % counter style
\subtablenumstyle{blank}                         % counter style

\eqnseriesstyle{alphabetic}                      % sub-counter style for series
\figseriesstyle{alphabetic}                      % sub-counter style for series
\tblseriesstyle{alphabetic}                      % sub-counter style for series

\def\puteqnformat{\hbox{%                        % equation number format
     \ifblank\chapternumstyle\then\else\numstyle\chapternum.\fi
     \ifblank\sectionnumstyle\then\else\numstyle\sectionnum.\fi
     \ifblank\subsectionnumstyle\then\else\numstyle\subsectionnum.\fi
     \numstyle\equationnum
     \numstyle\subequationnum}}
\def\putfigformat{\hbox{%                        % figure number format
     \ifblank\chapternumstyle\then\else\numstyle\chapternum.\fi
     \ifblank\sectionnumstyle\then\else\numstyle\sectionnum.\fi
     \ifblank\subsectionnumstyle\then\else\numstyle\subsectionnum.\fi
     \numstyle\figurenum
     \numstyle\subfigurenum}}
\def\puttblformat{\hbox{%                        % table number format
     \ifblank\chapternumstyle\then\else\numstyle\chapternum.\fi
     \ifblank\sectionnumstyle\then\else\numstyle\sectionnum.\fi
     \ifblank\subsectionnumstyle\then\else\numstyle\subsectionnum.\fi
     \numstyle\tablenum
     \numstyle\subtablenum}}

%************** Reference numbering *************************

\referencestyle{sequential}                      % referencing method
\referencenumstyle{arabic}                       % counter style
\def\putrefformat{\numstyle\referencenum}        % format of reference citation
\def\referencenumformat{\numstyle\referencenum.} % format of number in list
\def\putreferenceformat{%                        % paragraphing of list
     \everypar={\hangindent=1em \hangafter=1 }%
     \def\\{\hfil\break\null\hskip-1em \ignorespaces}%
     \leftskip=\refnumindent\parindent=0pt \interlinepenalty=1000 }

%************** Font initialization *************************

\normalsize

%*****************************************************************************

\def\fmtversion{2.6M (June 1992)}

\catcode`\@=12
% ------------------ End of jytex.tex -----------------

%\input jytex.tex   % available from hep-th
\typesize=10pt \magnification=1200 \baselineskip17truept
%\baselineskip25truept
\footnotenumstyle{arabic} \hsize=6truein\vsize=8.5truein
\input epsf
%draft
%\leftmargin=1.25in
%\oddleftmargin=.5in
%\evenleftmargin=1.5in
\sectionnumstyle{blank}
\chapternumstyle{blank}
\chapternum=1
\sectionnum=1
\pagenum=0
%\referencestyle{preordered}
% title style follows

\def\begintitle{\pagenumstyle{blank}\parindent=0pt
\begin{narrow}[0.4in]}
\def\endtitle{\end{narrow}\newpage\pagenumstyle{arabic}}

% exercise style follows

\def\beginexercise{\vskip 20truept\parindent=0pt\begin{narrow}[10
truept]}
\def\endexercise{\vskip 10truept\end{narrow}}

% **************    my jyTeX abbreviations   *****************

\def\eql#1{\eqno\eqnlabel{#1}}
\def\ref{\reference}
\def\peq{\puteqn}
\def\pref{\putref}

\def\mgn{\marginnote}
\def\bex{\begin{exercise}}
\def\eex{\end{exercise}}

% *********************** My definitions ************************

\font\open=msbm10 %scaled\magstep1 % For VAX. Borde p195.

 %scaled\magstep1 % For VAX. Borde p195.
%\font\open=msym10 %scaled\magstep1 % For Arbortxt on PC
%\font\opens=msym8 %scaled\magstep1 % For Arbortxt on PC
\font\goth=eufm10  % For Arbortxt on PC, and VAX. Borde p199

\def\StretchRtArr#1{{\count255=0\loop\relbar\joinrel\advance\count255 by1
\ifnum\count255<#1\repeat\rightarrow}}
\def\StretchLtArr#1{\,{\leftarrow\!\!\count255=0\loop\relbar
\joinrel\advance\count255 by1\ifnum\count255<#1\repeat}}

\def\StretchLRtArr#1{\,{\leftarrow\!\!\count255=0\loop\relbar\joinrel\advance
\count255 by1\ifnum\count255<#1\repeat\rightarrow\,\,}}

\def\mbox#1{{\leavevmode\hbox{#1}}}

\def\hspace#1{{\phantom{\mbox#1}}}
\def\oR{\mbox{\open\char82}}

\def\oZ{\mbox{\open\char90}}

\def\gE{\mbox{{\goth\char69}}}

\def\gS{\mbox{{\goth\char83}}}

\def\al{\alpha}
 %in jyTeX
 %in jyTeX
 %in jyTeX
 %in jyTeX
 %in jyTeX
 %in jyTeX
 %in jyTeX
 %in jyTeX
 %in jyTeX
 %in jyTeX
 %in jyTeX
 %in jyTeX
 %in jyTeX
% in jyTeX
% in jyTeX
% in jyTeX
\def\bom{{\bmit\omega}}% in jyTeX
% in jyTeX

\def\de{\delta}
\def\Ga{\Gamma}

\def\la{\lambda}

\def\om{\omega}
\def\Om{\Omega}

\def\si{\sigma}
\def\Si{\Sigma}
\def\th{\theta}

\def\De{\Delta}

\def\caC{{\cal C}}

\def\caS{{\cal S}}

\def\Real{{\rm Re\,}}

\def\sc{{\rm sc }}

\def\zf{$\zeta$--function}

     % Newline

\def\frac#1/#2{\leavevmode\kern.1em
\raise.5ex\hbox{\the\scriptfont0 #1}\kern-.1em/\kern-.15em
\lower.25ex\hbox{\the\scriptfont0 #2}}
\def\sfrac#1/#2{\leavevmode\kern.1em
\raise.5ex\hbox{\the\scriptscriptfont0 #1}\kern-.1em/\kern-.15em
\lower.25ex\hbox{\the\scriptscriptfont0 #2}}

\def\gtorder{\mathrel{\raise.3ex\hbox{$>$}\mkern-14mu
             \lower0.6ex\hbox{$\sim$}}}
\def\ltorder{\mathrel{\raise.3ex\hbox{$<$}\mkern-14mu
             \lower0.6ex\hbox{$\sim$}}}

\def\semidirprod{\rlap{\ss C}\raise1pt\hbox{$\mkern.75mu\times$}}
\def\for{\lower6pt\hbox{$\Big|$}}
\def\fish{\kern-.25em{\phantom{abcde}\over \phantom{abcde}}\kern-.25em}

 %triple
%dot
 %double
%dot
 %double dot
%for small #1

\def\boxit#1{\vbox{\hrule\hbox{\vrule\kern3pt
        \vbox{\kern3pt#1\kern3pt}\kern3pt\vrule}\hrule}}
\def\dalemb#1#2{{\vbox{\hrule height .#2pt
        \hbox{\vrule width.#2pt height#1pt \kern#1pt \vrule
                width.#2pt} \hrule height.#2pt}}}

        %double stroke
\def\frac#1#2{{{#1}\over{#2}}}
 %lower covariant deriv.
 %upper covariant deriv.
 %lower covariant deriv semicolon.
    %lower ordinary  deriv.
    %lower ordinary  deriv comma.

\def\noin{\noindent}

      %Connection
    %Connection'
\def\comb#1#2{{\left(#1\atop#2\right)}}

\def\eg{{\it e.g.}}
\def\ie{{\it i.e. }}
\def\cf{{\it cf }}
\def\pa{\partial}

\def\av#1{\langle#1\rangle} %gives average <#1>
 %gives thermal average <<#1>>
   %gives bracket <#1|#2>
   %gives comma bracket <#1,#2>
 %gives round bracket (#1,#2)
 %gives round bracket (#1,|#2)
 %gives big bracket <#1|#2>
  %gives
%matrix element <#1|#2|#3>

%gives reduced matrix element
%<#1||#2||#3>

\def\3j#1#2#3#4#5#6{\left\lgroup\matrix{#1&#2&#3\cr#4&#5&#6\cr}
\right\rgroup}

\def\m?{\mgn{?}}
% KK's defs

\def\pa{\partial}

\def\beq{\begin{eqnarray}}
\def\eeq{\end{eqnarray}}

\def\nsl{\nabla\!\!\!\! / }

%  *******************  Journal refs **********************

\def\cmp#1#2#3{{\it Comm. Math. Phys.} {\bf {#1}} ({#2}) #3}

\def\jmp#1#2#3{{\it J. Math. Phys.} {\bf {#1}} ({#2}) #3}
\def\jpa#1#2#3{{\it J. Phys.} {\bf A{#1}} ({#2}) #3}

\def\np#1#2#3{{\it Nucl. Phys.} {\bf B{#1}} ({#2}) #3}

\def\prD#1#2#3{{\it Phys. Rev.} {\bf D{#1}} ({#2}) #3}

\def\am#1#2#3{{\it Acta Mathematica} {\bf {#1}} ({#2}) #3}

\def\jpamt#1#2#3{{\it J. Phys.A:Math.Theor.} {\bf{#1}} ({#2}) #3}

\def\pams#1#2#3{{\it Proc. Am. Math. Soc.} {\bf {#1}} ({#2}) #3}

\def\plb#1#2#3{{\it Phys. Letts.} {\bf {B#1}} ({#2}) #3}

\def\qjm#1#2#3{{\it Quart. J. Math.} {\bf {#1}} ({#2}) #3}

% *******************   Main text *********************
%\draft
\begin{title}
\vglue 0.5truein
%\righttext {MUTP/96/23}
%\righttext{hep-th/96}
\vskip15truept
%\leftline{\today}
%\vskip 30truept
\centertext {\Bigfonts \bf R\'enyi entropy for monodromy defects of higher } \vskip7truept
\vskip10truept\centertext{\Bigfonts \bf derivative free fields on even--dimensional spheres} \vskip17truept
\centertext{\Bigfonts \bf }
 \vskip 20truept
\centertext{J.S.Dowker\footnote{dowkeruk@yahoo.co.uk}} \vskip
7truept \centertext{\it Theory Group,} \centertext{\it School of Physics and Astronomy,}
\centertext{\it The University of Manchester,} \centertext{\it Manchester, England} \vskip
7truept \centertext{}
\vskip 7truept

\vskip40truept
\begin{narrow}
Explicit polynomial forms for R\'enyi and entanglement entropies are given on even --dimensional spheres which possess a spherical codimension--2 U(1) monodromy defect. Free scalar and Dirac fields are treated and higher-derivative propagation operators employed. The central charge, $C_T$, is also calculated.

The results are compared with existing  ones for a planar defect. It is shown that these can be obtained from the values in the present paper.

 \end{narrow}
\vskip 5truept
%\righttext {August 1996}
\vskip 60truept
%\righttext{Typeset in \jyTeX}
\vfil
\end{title}
\pagenum=0
\newpage

\section{\bf 1. Introduction.}

This paper is a continuation of an earlier one, [\pref{dowresm}], which was concerned with a free scalar field propagating according to the Penrose--Yamabe conformally covariant Laplacian, $Y_d$. It is of some technical interest to extend the analysis to spinors and also to the GJMS--type higher derivative operators which, conveniently, factorise on the sphere, S$^d$,(as well as some other, related, manifolds) leading to amenable spectral problems on S$^d$ and its factors such as conical deformations.

The availability of higher derivative expressions provides useful  information, as demonstrated in Beccaria and Tseytlin, [\pref{BandT}], for example.

For fields which are simply periodic on the sphere an account of the formalism and results is given in [\pref{dowgjmsren}] . This is here generalised to a (complex) field that suffers a phase change due to an Aharonov--Bohm type flux through the sphere constituting, in present terminology, a  U(1) monodromy defect. The previous paper, [\pref{dowresm}], gives some history of the topic with a few relevant references, old and new.

The quantities of interest are the free energy (effective action) and its various derivatives such as the entanglement entropy. The aim is to provide explicit formulae for as many dimensions, $d$, as feasible, which, despite being for a somewhat restricted situation,  should be helpful for comparison and checking purposes.

\section{\bf 2. The operators}
 For completeness, it is necessary to give a few details, firstly of the propagating  operators which here have the  Branson structure, on the sphere,
  $$
  \Om_k(d)={\Ga(B+k+1/2)\over\Ga(B-k+1/2)}\,,
  \eql{intertw}
  $$
where $k$ is a  parameter, either an integer or a half--integer, to begin with. The (pseudo) operator $B$ depends on the field theory. For scalars $B= \sqrt{Y_d+1/4}$ and for Dirac spinors $B= (\nsl^{\,2})^{1/2}=|\nsl|$.

When $k$ is integral or half integral, the form (\peq{intertw}) reduces to a finite product
of operators. To accord with the ordinary field case, $k$ is an integer for integer spin and
a half--integer for half--integer spin. These give, respectively, an even or an odd number of
factors. More generally, $k$ has a conformal field theory significance in terms of a scaling
dimension, \cf\ [\pref{Diaz,DandD,KPS2}].

The operator (\peq{intertw}) for integer $k$ reads,
  $$\eqalign{
      \Om_k(d)=&\prod_{j=0}^{k-1} \big(B^2-(j+1/2)^2\big)
      =\prod_{j=0}^{k-1} \big(B-(j+1/2)\big)\big(B+(j+1/2)\big)\,,\cr
      }
      \eql{op}
  $$
expressed as a product of $2k$ linear factors.

 \section{\bf3. The spectra}
As in [\pref{dowresm}], there is a choice of methods. One involves only the eigenvalue structure, and the other the propagating Green function. Here I will concentrate on the former as being more directly global, avoiding construction of complicated local field theory quantities, like  the energy--momentum tensor, and their integration.

It was shown in [\pref{Dowchargedrenyi}] that the spectrum on a $q$--deformed sphere for a charged scalar undergoing a phase change of $2\pi\de$ on circling the flux is the union of two sets of eigenvalues,
$$\eqalign{
&
   \bigg\{\big({d-1\over2}+ q|\de|+{\bom}\cdot{\bf n}\big)^2-\al^2\bigg\}\cup
    \bigg\{\big({d-1\over2}+ q-q|\de|+{\bom}\cdot{\bf n}\big)^2-\al^2\bigg\}\,,
   }
   \eql{eigsets}
  $$
where $\bom$ is the set of $d$ integers, $(q,{\bf1}_{d-1})$, but can be considered as
real non--negative and referred to as the {\it parameters}. $\bf n$ is a set of $d$
integers, $(n_1,n_2,\ldots,n_d)$ each ranging from 0 to $\infty$. If the propagating
operator is conformal, \ie equals $Y_d$, then $\al=1/2$. Degeneracies arise simply from coincidences between the eigenvalues as displayed in (\peq{eigsets}) and need not be specifically introduced being encoded in the sums over ${\bf n}$.

The Dirac eigenvalues are, likewise,
$$\eqalign{
&
   \bigg\{\big({d-1+q\over2}+ q|\de|+{\bom}\cdot{\bf n}\big)^2-\al^2\bigg\}\cup
    \bigg\{\big({d-1+q\over2}-q|\de|+{\bom}\cdot{\bf n}\big)^2-\al^2\bigg\}\,,
   }
   \eql{eigsetsd}
  $$
where $\al=0$ for conformal invariance \ie propagating operator $\nsl$. (The eigenvalues are then for the square, $\nsl^{\,2}$.)

These spectra show that the eigenvalues of the product $\Om_k$, (\peq{op}), involve the quantities,
  $$
  \la({\bf n},a,q,\de)^2-\al^2,
  $$
where
$$
\la({\bf n},a,q,\de)\equiv (a+{\bf n.}\bom)\,,\quad {\bom}=(q,1,\ldots,1)\,,
  \eql{lam}
$$
associated with  each factor ($2k$ in number) labelled by $j$ with, explicitly, $\al\equiv\al_j=j+1/2$, $j=0,1,\ldots k-1$. There
are two values of the parameter $a$ to take into account for each spin, and it is useful to write them out,
   $$
        a^S_1={d-1\over2}+q\de\,,\quad a^S_2={d-1\over2}+q(1-\de)\,,
        \eql{ays}
   $$
for scalars, and,
  $$
        a^F_1={d-1+q\over2}+q\de\,,\quad a^F_2={d-1+q\over2}-q\de\,,
   $$
 for spinors. $a^F$ are got from $a^S$ by the replacement $\de\to\de+1/2$. The shift of 1/2 reflects the spinor sign change encounted on a complete circuit of the flux. Note that periodicity is then with respect to the unit cell $-1/2\le\de\le1/2$.\footnote{ This can be seen by remarking that $1/2+\de$ for spinors plays the role that $\de$ plays for scalars \ie it is $1/2+\de$ that has unit cell $0$ to $1$. Periodicity then {\it implies} that $1-\de$ is equivalent to $\de$.}

\section{\bf 4. Conformal anomaly}.

In even dimensions, the conformal anomaly provides the universal coefficient of the logarithmic UV--divergent term in the effective action and equals the value of the propagating \zf, here denoted by $Z_k(s,d,q,\de)$, at $s=0$, up to a multiplicative constant (and assuming no zero modes).

The evaluation of $Z_k(0,d,q,\de)$ is described in [\pref{DowGJMS}] based on [\pref{Dowcmp}]. It relies on the linear factorisation of the product $\Om_k$ as shown in (\peq{op}). For convenience, I outline the procedure.

The \zf\ corresponding to the product (\peq{op}) is given by,
$$\eqalign{
  Z_k(s,d,q,\de)&=\caS\sum_i\,
  \sum_{\bf n=0}^\infty \prod_{j=0}^{k-1}{1\over\big(\la({\bf n},a_i-\al_j,q)\big)^s\big(\la({\bf n},a_i+\al_j,q)\big)^s}\cr
  \,,\cr
  }
  \eql{zedk}
  $$
where $\caS$ is the field dimension. For complex scalars $\caS=2$ and for Dirac spinors,
$\caS=2^{d/2+1}$.

Although the even product, (\peq{op}), applies, as it stands, just to scalar fields, it can still be employed for spinors by continuing $k$ to a half--integer. (See (\peq{intertw}).)

The \zf\ of a linear factor in (\peq{zedk}) can be recognised as a Barnes \zf,  each one contributing additively to the value at $s=0$, $Z_k(0,d,q,\de)$, which, when multiplied by $k$, gives the required conformal anomaly,
  $$
  \caC(d,q,\de,k)=(-1)^{2s}k\,Z_k(0,d,q,\de)\,\quad (s={\rm spin})\,.
  $$
The value of the Barnes \zf\ at 0 is given by a generalised Bernoulli polynomial, $B^{(d)}_\nu(*)$ which is easily accessible.

For each pair of linear factors (\ie value of $j$ in (\peq{zedk})) there are four values of the argument of the polynomials and so,
       $$
      \caC(d,\de,q,k)={(-1)^{2s}\caS\over 2d!q}\sum_{\pm}\sum_{j=0}^{k-1}
      \bigg( B^{(d)}_d\big(a_1\pm\al_j\mid q,{\bf1}_{d-1}\big)+ B^{(d)}_d\big(a_2\pm\al_j\mid q,{\bf1}_{d-1}\big)\bigg)\,.
      \eql{cak}
      $$

Symmetry properties of the $B^{(d)}_\nu$ allow some simplification. For example for even $\nu$, \footnote{ For odd $\nu$ there is a sign change and the expressions would cancel to give zero conformal anomaly in odd dimensions, as is correct.}
      $$\eqalign{
         B^{(d)}_{\nu}\big(a^S_2\pm\al_j\mid q,{\bf1}_{d-1}\big)&=
         B^{(d)}_\nu\big (d-1)/2+q-a^S_2\mp\al_j\mid q,{\bf1}_{d-1}\big)\cr
         &=B^{(d)}_\nu\big(a^S_1\mp\al_j\mid q,{\bf1}_{d-1}\big)\,,
         }
      $$
using the expressions for $a_1^S$ and $a_2^S$ in (\peq{ays}). The same result holds for the spinor arguments. Hence the second term in (\peq{cak}) can be expunged with  a compensating  overall factor of 2.

Furthermore, the GJMS sum over $j$ can be performed to give the more compact expression, with a bulk, $(d+1)$--dimensional  holographic appearance,
  $$\eqalign{
 & \caC(d,q,\de,k)=\cr
  &{(-1)^{2s}\caS\over (d+1)!q}\bigg(B^{(d+1)}_{d+1}\big(a_1+k+1/2\mid q,{\bf1}_d\big)-B^{(d+1)}_{d+1}\big(a_1-k+1/2\mid q,{\bf1}_d\big)\bigg)\,.\cr
  }
  \eql{cak1}
  $$

It is to be noted that $k$ can now be continued off the integers, in particular to $k=1/2$ which delivers the conventional Dirac value with flux $\delta$ {\it if} $\de$ in (\peq{cak1}) is replaced by $\de+1/2$. This is a useful calculational shortcut.\footnote{  Of course the odd product (1) obtained when $k$ is half--integral can be used. This would mean extra work, but can be helpful as a check. As functions of $k$, the quantities can be regarded as giving a sort of interpolation between scalar and spinor, up to a sign and spin factor.}

The evaluation of  the polynomials is straightforward and rapid by nachine  but, as the expressions become unwieldy, I give only a single example which exhibits the general structure. Omitting the $(-1)^{2s}\caS$ factor, there results, in 4 dimensions,
$$
\caC(4,q,\de,k)=k{11-20 k^2+6 k^4\over q}-k q^3 (30 \si^2-1)-10k q (1-k^2) (6 \si-1)\,,
$$
where $\si=\de(1-\de)$ is a more convenient variable.

I list below a few examples of the defect contribution, for the full sphere,  to the conformal anomaly for a Dirac field, defined by,
$$\De\caC(d,1,\de,k)\equiv\caC(d,1,1/2+\de,k)-\caC(d,1,1/2,k)\,,\quad k\in\oZ_{n+1/2}\,,
$$
again omitting $\caS$.

For a standard Dirac field,
$$\eqalign{
\De\caC(4,1,\de,1/2)&={1\over24} \de^2 (2 - \de^2),\cr
\De\caC(6,1,\de,1/2)&=-{1\over1440}\de^2 (24 - 15 \de^2 + 2 \de^4),\cr
\De\caC(8,1,\de,1/2) &={1\over120960}\de^2 (432 -294 \de^2 + 56 \de^4 - 3 \de^6),\cr
\De\caC(10,1,\de,1/2)&= -{1\over7257600}\de^2 (5760 - 4100 \de^2 + 910 \de^4 - 75 \de^6 + 2 \de^8)\,,
   }
   \eql{dcsp1}
$$
and for a cubic Dirac field,
$$\eqalign{
\De\caC(4,1,\de,3/2)&=-{1\over8} \de^2 (2 + \de^2), \cr
\De\caC(6,1,\de,3/2)&={1\over480} \de^2 (16 - 5 \de^2 - 2 \de^4), \cr
\De\caC(8,1,\de,3/2)&=-{1\over13440}\de^2 (80 - 42 \de^2 + \de^6), \cr
\De\caC(10,1,\de,3/2)&= {1\over2419200}\de^2 (2880 - 1780 \de^2 + 210 \de^4 + 15 \de^6 -2 \de^8)\,.
  }
  \eql{dcsp3}
$$

\section{\bf5. The term proportional to q}
This section and the next contain observations on aspects  of the form of the conformal anomaly not needed for the main calculation which continues in section 7.

For standard conformally invariant scalar fields, the term proportional to $q$ in the conformal anomaly vanishes. For higher derivative propagation, and no flux,  it is a polynomial in $k$ that vanishes when $k$ is integral and satisfies the GJMS existence (`subcritical') condition $k< d/2$, as described in [\pref{dowgjmsren}]. Calculation shows that the same holds true in the presence of flux, the coefficient being,
  $$
        {(-1)^{d/2-1}\over (d-1)(d-2)!}\,\big({1\over6}-\si\big)k(k^2-1^2)(k^2-2^2)\ldots\big(k^2-(d/2-1)^2)\,.
        \eql{coef}
  $$
The derivation of this formula is oulined in an appendix.

  It is interesting to note that the coefficient, (\peq{coef}), is zero for all $d$ and $k$ when $\si=1/6$ \ie a scalar flux of $\de=(\sqrt3-1)/2\sqrt3\approx 0.2113249$ and a Dirac flux of $1/\sqrt{12}\approx0.288675$. I do not know what, if anything, this signifies.

  \section{\bf 6. The small $q$ limit}

As a check of the numerics, and as an interesting fact, the coefficient  of the $1/q$ term in the  conformal anomaly is found to equal minus twice the vacuum energy on the Einstein universe, $\oR\times$S$^{d-1}$. An analytical proof of this can be given that relies on simple properties of the Bernoulli polynomials. Indeed, the coefficient of the $1/q$ term is a polynomial in $k$ for which a general Bernoulli formula is given in [\pref{dowqretspin}] realised as the vacuum energy of a Rac $k$-lineton on the Einstein universe.

The limit $q\to0$ corresponds to an infinitely sheeted cover of the sphere, S$^d$,  and is independent of $\de$, which is the flux through the entire covering. A different result is obtained if the flux, $\mu$, through one sheet is required to remain constant by setting
$\mu=q\de$. This can be thought of as a Euclidean chemical potential. The earlier calculations, [\pref{B,Dowchargedrenyi}] then become more relevant.\footnote{ Standard fields only were treated here.}

\section{\bf7. Other quantities. Entropies and conformal charge.}

From the effective action (`free energy'), \ie here just the conformal anomaly, other quantities of interest can be derived, such as the R\'enyi and entanglement entropies and various central charges. I will not repeat the standard definitions.

All these quantities are polynomial in $q,k$ and $\de$. I will generally display only for specific values of $k$. For example, the R\'enyi entropy, $\gS(d,q,\de,k)$, for standard Dirac is ($q=1/n$),
$$
\gS(4,q,\de,1/2)=
{1 + q\over5760} \big(37 - 480 \de^2 + 240 \de^4 +q^2 (7 - 120 \de^2 + 240 \de^4)\big)\,.
$$

I list some entanglement entropies, $\gE(d,\de,k)$, of a standard Dirac spinor ($k=1/2$ and flux $\de$) on the full sphere (the physical spin factor has now been included),
$$\eqalign{
\gE(4,\de,1/2)&={11\over90}-{1\over3} \de^2 (5 - 4 \de^2),\cr
\gE(6,\de,1/2)&=- {191\over3780} +{1\over90} \de^2 (64 - 65 \de^2 +12 \de^4),\cr
   \gE(8,\de,1/2)&={2497\over113400}-{1\over3780}\de^2 (1188 - 1323 \de^2 +350 \de^4-24 \de^6),\cr
   \gE(10,\de,1/2)&=-{14797\over1496880}+{8\over14175} \de^2 (16128 - 18860 \de^2 + 5824 \de^4 - 615 \de^6 + 20 \de^8)\,.
   }
   \eql{dente}
$$

The constant term is minus the usual conformal anomaly. The remainder is the effect of the monodromy.

 The energy--momentum two--point central charge, $C_T$, follows from the second derivative (with respect to $1/q$) of the free energy (conformal anomaly) using the Perlmutter factor. For free scalar and Dirac fields, and no flux, general formulae exist as functions of the dimension $d$ and the derivative order $2k$.

 For standard ($k=1$) scalar fields some expressions for $C_T$ were given in [\pref{dowresm}]. Here I extend these to higher derivatives and to spinors. The notation is $C_T(d,\de,k)$.

 For the scalar, Paneitz--Tseytlin four--derivative operator,\mgn{CHECK times2 for complex}
 $$\eqalign{
 C_T(4,\de,2)&=-{64\over3} + 160\si (1-\si), \cr
 C_T(6,\de,2)&=-12 +42\si^2 (9-2 \si),\cr
 C_T(8,\de,2)&= -{64\over7} +{16\over3} \si^2 (46 + 26\si - 3 \si^2),\cr
 C_T(10,\de,2)&=-{70\over9}+{55\over216} \si^2\big (776-6\si^3+69\si^2+640\si\big)\,.
   }
 $$

As in the standard case, as the flux is turned on, $C_T$ changes sign, here from negative to positive at a root, $\si_0(d)$,  in the unit cell ($0\le\de\le1$, \ie $0\le\si\le1/4$) of the polynomial. Typically, $\si_0(8)\approx0.18391467$. The corresponding flux is $\de_0(8)\approx0.24292933$.

For a standard ($k=1/2$) Dirac field, omitting the spin factor $\caS$, ($\de$ is the flux),
$$\eqalign{
C_T(4,\de,1/2)&=2 - 30 \de^2 + 40 \de^4, \cr
C_T(6,\de,1/2)&=3 - {21\over4} \de^2 (\de^2-3) ( 4 \de^2-3), \cr
C_T(8,\de,1/2)&=4-{1\over3}( 193 \de^2 + 350 \de^4 -133 \de^6 + 12 \de^8), \cr
C_T(10,\de,1/2)&=5 - {55\over576} \de^2 (852 - 1615 \de^2 + 714 \de^4 - 99 \de^6 + 4 \de^8)\,.
 }
$$
$C_T$ changes sign at $\de=\de_0(d)$, with, typically, $\de_0(10)\approx\pm0.2657463$.
The defect contribution is obtained by dropping the first (standard) term, $d/2$.

For cubic Dirac ($\nsl^{\,3}$),
$$\eqalign{
C_T(4,\de,3/2)&=-{2\over3} - 10 \de^2 +120 \de^4,\cr
C_T(6,\de,3/2)& =-{18\over5} + {63\over4}\de^2 (\de^2 + 4 \de^4-3), \cr
C_T(8,\de,3/2)&=-{36\over7} + 77 \de^2 -100 \de^4 - 13 \de^6 + 12 \de^8,\cr
 C_T(10,\de,3/2)&=-{115\over18}  +{55\over1728} \de^2 (5065 \de^2 -
 1022 \de^4 - 219 \de^6 + 36 \de^8-3140)
    }
$$
and, \eg, $\de_0(10)\approx\pm0.26872496$.

It is interesting to note that, for higher derivatives, as a function of a {\it complex} flux, $C_T$ has zeros on the imaginary axis for fermions while for bosons the real part is shifted to 1/2. For example, for the Paneitz case, $C_T(10,\de,2)$, the roots are $\approx1/2\pm4.19796i$ while for $k=3$ there are four roots, $\approx1/2\pm2.0606 i$ and $\approx1/2\pm
7.41175 i$. in the complex unit cell , $0\le \Real\de\le1$. \footnote{ At an earlier stage, a similar behaviour occurs for the basic free energy. The root imaginary part would correspond to a real chemical potential while the real part turns bosons into fermions.} It is amusing to analyse the root structure a little further.

If $k$ is extended to a real number, the central charge, for example, can vanish at more than two values of the flux. The graph below displays the variation, for dimension $d=6$, of scalar $C_T$ against $\de$ for $k=1.2$, showing four real roots.

\vskip  5truept

\epsfbox{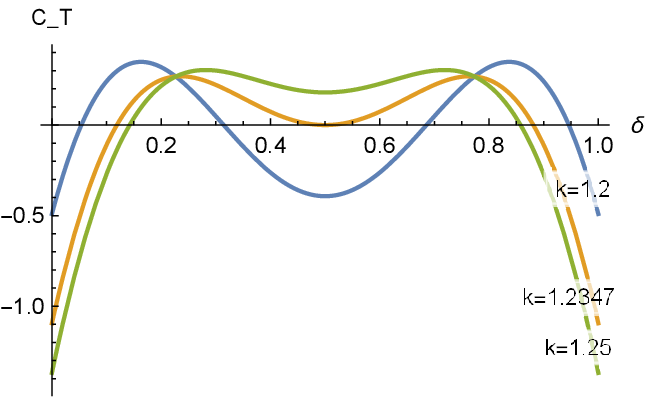}

As $k$ increases towards $2$, say, the two inner roots come together until they coincide at $\de=1/2$ ($\si=1/4$), the value of $k$ being approxinately $1.2346900$.
Thereafter, as shown \eg\ in the graph for $k=1.25$,  these two roots have disappeared to become  complex conjugate ones . As $k$ increases further, this process repeats itself with complex roots, of real part 1/2, being created as $k$ passes through certain non--integer values. For each $d$, the number of these $k$--values is finite, the total number, including 0 and negative values, is $d-1$.

The behaviour in the Dirac case is similar except that the vertical axis is shifted by 1/2 to the right.

A further fact worth remarking is that the periodic extension of $C_T$ displays cusps at the boundary of the unit cell.

\section{\bf 8. Comparisons. A tale of two defects}
A prediction, when $d=6$, is given in [\pref{C+}] for the effect of a codimension 2 monodromy defect on the Dirac entanglement entropy which I copy here, dropping the extra singular mode,\footnote{ According to the authors of [\pref{C+}] there is a substitutional error in [\pref{C+}] and I have recalculated the expression using the forms of $a_\Si$ and $d_2$ as given in this reference.}
$$
  - {1\over90}  \de^2 (16-5\de^2)\,.
$$

This differs from the value found here which is, from (\peq{dente}),
$$
{1\over90} \de^2 (64 - 65 \de^2 +12 \de^4)\,.
\eql{sc}
$$
The sign difference is a consequence of conventions but the coefficients differ. In particular, (\peq{sc}) is a higher degree polynomial.
As noted in [\pref{dowresm}], there is also  a  similar disparity in the  scalar field quantities.

The origin of this difference can be traced to the geometric positioning, relative to the entangling surface, of the particular defects involved. In fact, the two defects could be thought of as being orthogonal. In the case discussed in the present paper, the defect is spherical and coincident with the entangling surface while that in [\pref{C+}] is planar and cuts through the (same) entangling surface which is centred on the defect.

The analytical consequences of this are to be found in a term in the expression for the entropy derived in [\pref{KNSW}] via an expansion of the replica formula about $n=1$. The derivative part of this is the term in question and consists of an integral over the hyperbolic cylinder of a vacuum--averaged, ambient conformal energy density, $\av{T^\tau_\tau}$, which is related by conformal transformation to the vacuum average of the energy--$\!\!$ momentum tensor in flat space. From its general form, given in [\pref{KNSW}], the two energy densities, corresponding to the two defect geometries, are related by a factor of $-(d-1)$.\footnote{ I am grateful to the authors of [\pref{C+}] for explaining this particular point and also the geometry involved in the two cases.} In particular, $\av{T^\tau_\tau}$ equals (to factors of $\sqrt g$ and anomalies) the conventional, flat space energy density, $\av{T^0_0}$, for the planar defect in [\pref{C+}] but equals the angular, or Rindler, average, $\av{T^\th_\th}=-(d-1)\av{T^0_0}$ for the spherical defect employed here. The other part of the expression for the defect  entanglement entropy is the difference in free energies, and is the same for both defects.

These facts allow the entanglement entropy for the planar defect to be obtained  easily by simple algebra from that for the spherical one computed in this paper.

I find that, for ordinary scalar fields the planar defect entanglement entropy difference, $\De\gE_P(d,\de,k)$, is, in my sign conventions,
$$\eqalign{
\De\gE_P(4,\de,1)=&\,0,\cr
 \De\gE_P(6,\de,1)=&{\si^2\over180}, \cr
\De\gE_P(8,\de,1)= &-{\si^2 (9 +
   2 \si)\over7560},\cr 
 \De\gE_P(10,\de,1)=  & {\si^2 (216 + 72 \si +
    5 \si^2)\over907200}\,. \cr
    }
$$
The first two values are given in [\pref{B+,C+}] which have thus been confirmed, and extended, using a distinct, global technique.

For the Paneitz case,
$$\eqalign{
\De\gE_P(4,\de,2)=&{4\si\over3},\cr
 \De\gE_P(6,\de,2)=&-{\si^2\over18}, \cr
\De\gE_P(8,\de,2)= &{\si^2 (6 +
   \si)\over945}\cr
  \De\gE_P(10,\de,2) =&-{\si^2 (444 + 128 \si + 5 \si^2)\over453600}\,.
    }
$$

For ordinary spinors (flux $=\de$ and omitting $\caS$),
$$\eqalign{
\De\gE_P(4,\de,1/2)=&-{\de^2\over24},\cr
\De\gE_P(6,\de,1/2)=& {\de^2 (16 - 5 \de^2)\over1440},\cr
\De\gE_P(8,\de,1/2)=& -{\de^2 (324 - 147 \de^2 + 14 \de^4)\over120960},\cr
 \De\gE_P(10,\de,1/2)=& {\de^2 (4608 - 2460 \de^2 + 364 \de^4 - 15 \de^6)\over7257600}\,.
}
$$
Again, the first two values confirm those in [\pref{B+,C+}].

For cubic Dirac,
$$\eqalign{
\De\gE_P(4,\de,3/2)=&{5 \de^2\over24},\cr
\De\gE_P(6,\de,3/2)=& -{\de^2 (4-\de^2)\over160},\cr
\De\gE_P(8,\de,3/2)=& {\de^2 (188 - 73 \de^2 +
   2 \de^4)\over40320}, \cr
   \De\gE_P(10,\de,3/2)=&-{\de^2 (7056 - 3400 \de^2 + 308 \de^4 + 5 \de^6)\over7257600}\,.
}
$$

Finally, in tis section, I remark that the defect considered in the present paper is the same as that investigated in [\pref{GHJK}].

\begin{ignore}
\section{\bf8. Central charge closed formulae}

For higher derivative scalars, an expression for $C_T$, holding for all dimensions  and derivative orders, was conjectured by Osborn and Stergiou, [\pref{OandS}]. In the same paper, as an addendum, a proof was given based on an existing flat--space four--point function. A CFT proof was also given by Gliozzi {\it et al}, [\pref{GGPW}], who refer to the fields as `generalised free fields'.

In [], a direct spherical  spectral derivation was given using the objects appearing in the present work. The Dirac expression was also obtained. I have not seen this elsewhere.
The proof used the  Perlmutter  factor.

Reference [\pref{GGPW}] defines a {\it total central charge} as the sum of the central charges for $k=1,2,\dots$ \ie including the values above the GJMS limit ($d/2$) which is where the numerator begins to have poles. Assuming dimensional regularisation, the infinite sum is performed by machine.
\end{ignore}

\section{\bf8. Comments and conclusion}

The results of a machine evaluation of several quantities, based on the conformal anomaly, have been presented for free fields satisfying higher--derivative propagation equations on even--dimensional spheres in the presence of a spherical monodromy defect.

Comparison with existing results highlights the differing geometrical constructions of the defects and their analytical consequences.

The variation of a central charge with the monodromy flux is investigated.

In a further communication, I will discuss the case of $p$--forms which have a more complicated field theory. The analysis will also be extended to odd dimensions, part of which already exists in [\pref{dowmono}].

\section{\bf Acknowledgment}

I thank the authors of [\pref{C+}] for their very helpful communications.

\section{\bf Appendix. A little Bernoulli--ology}

In this appendix I derive the expression for the term proportional to $q$ in the conformal anomaly. It is given by
  $$
        {1\over2}   {\pa^2\over\pa q^2}\,q\, \caC(d,q.\de,k)\bigg|_{q=0}
        \eql{coeffq}
  $$
which means, from (\peq{cak1}) that one needs,
  $$
            {\pa^2\over\pa q^2}\, B^{d+1}_{d+1}\big(d/2+\de q\pm k\mid q,{\bf}1_{d}\big)\bigg|_{q=0}\,.
  $$

  The required $q$--polynomial  is provided by the standard formula (I give the general form, [\pref{Norlund1}]) ,
     $$
         B^{(n)}_\nu(x\mid\bom)=\sum_{s=0}^\nu\comb\nu s\,\om_1^sB_sB^{(n-1)}_{\nu-s}(x\mid\widehat{\bom})\,,
     $$
  where $\widehat\bom$ is the set $\bom$ omitting the first element, $\om_1$. For present application,  $n=d+1$, $\nu=d+1$ and $\om_1=q$ with $\widehat\bom={\bf1}_d$. This formula allows one to find the derivatives with respect to the $q$--dependence coming from the parameters $\bom$.

  The derivative with respect to the $q$--dependence in the argument, $x=d/2+\de q\pm k$, is given by the basic recursion,
     $$
             {\pa\over\pa x}\, B^{(n)}_\nu(x\mid\bom)=\nu B^{(n)}_{\nu-1}(x\mid\bom)\,.
             \eql{rec}
     $$

 Taking the second derivative produces four terms corresponding to the two sources of $q$--dependence. The derivatives with respect to the argument give factors  of $\de$ and simple algebra yields,
   $$
   {\pa^2\over\pa q^2}\,B^{(d+1)}_{d+1}(d/2+\de q\pm k\mid q,{\bf1}_d)\bigg|_{q=0}=
   d(d+1)(1/6-\si)B^{(d)}_{d-1}(d/2\pm k)\,,
   $$
  where the parameters ${\bf1}_d$ in the final Bernoulli polynomial have been conventionally omitted.

  Inserting the factors to give the conformal anomaly, (\peq{cak1}), gives for the coefficient  of the $q$ term,
    $$
    {1-6\si\over6(d-2)!(d-1)}B^{(d)}_{d-1}(d/2+k)\,,
    $$
 where (for even $d$) the antisymmetry of the Bernoulli polynomial in $k$ has been used.

 Finally, the product form, (this follows from (\peq{rec})),
    $$
        B^{(n+1)}_n(x)=(x-1)(x-2)\ldots(x-n)\,,
        \eql{fac}
    $$
converts the above expression into the one in section 5.

%\newpage
 \vglue 20truept
 \noin{\bf References.} \vskip5truept
\begin{putreferences}
\ref{dowqretspin}{Dowker,J.S. {\it Revivals and Casimir energy for a free Maxwell field
  (spin-1 singleton) on $R\times S^d$ for odd $d$}, arXiv:1605.01633.}
\ref{dowresm}{Dowker,J.S. {\it R\'enyi entropy for spherical monodromy of free scalar fields in even dimensions}, arXiv arXiv:2112.09031.}
 \ref{CandD}{Candelas,P. and Dowker,J.S. \prD{19}{1979}{2902}.}
\ref{KNSW}{Kobayashi,N., Nishioka,T., Sato,Y. and Watanabe,K. {\it Towards a C-theorem in defect CFT}, {\it JHEP} {\bf01} (2019) 170, arXiv 1810.06995.}
\ref{dowgjmsren}{Dowker,J.S. {\it R\'enyi entropy and $C_T$ for higher derivative free scalars and spinors on even spheres}, arXiv:1706.01369 .}
\ref{BandT}{Beccaria,M. and Tseytlin,A.A. {\it $C_T$ for higher derivative conformal
  fields and anomalies of (1,0) superconformal 6d theories}, arXiv:1705.00305.}
  \ref{GPW}{Guerrieri, A.L., Petkou, A. C. and Wen, C. {\it The free $\si$CFTs},
  arXiv:1604.07310.}
  \ref{GGPW}{Gliozzi,F., Guerrieri, A.L., Petkou, A.C. and Wen,C.
   {\it The analytic structure of conformal blocks and the
   generalized Wilson--Fisher fixed points}, {\it JHEP }1704 (2017) 056, arXiv:1702.03938.}
  \ref{YandZ}{Yankielowicz, S. and Zhou,Y. {\it Supersymmetric R\'enyi Entropy and
  Anomalies in Six--Dimensional (1,0) Superconformal Theories}, arXiv:1702.03518.}
  \ref{OandS}{Osborn.H. and Stergiou, A. {\it $C_T$ for Non--unitary CFTs in higher dimensions},
  {\it JHEP} {\bf06} (2016) 079, arXiv:1603.07307.}
  \ref{Perlmutter}{Perlmutter,E. {\it A universal feature of CFT R\'enyi entropy}
  {\it JHEP} {\bf03} (2014) 117. arXiv:1308.1083.}
   \ref{Norlund}{N\"orlund,N.E. {\it M\'emoire sur les polynomes de Bernoulli}, \am{43}{1922}{121}.}
        \ref{Lewin}{Lewin, L. {\it Polylogarithms and associated functions} (North--Holland,
        1981).}
        \ref{KandK3}{Koyama, S-Y. and Kurokawa, N. {\it Kummer's formula for multiple gamma functions}. {\it J. Ramanujan Math. Soc.} {\bf18} (2003) 87-107.}
        \ref{Kurokawa2}{Kurokawa,N. {\it Multilple zeta functions: An Example}, {\it Advanced Studies in Pure Mathematics} {\bf21} (1991) 219.}
       \ref{Kurokawa1}{Kurokawa, N. {\it Multiple sine functions and Selberg's zeta function},  {\it Proc.Jap.Acad}. {\bf67A} (1991) 61.}
      \ref{Dowcrit}{Dowker.J.S. {\it Conformal anomalies of higher derivative free critical $p$-forms on even spheres},arXiv:2007.13670.}
      \ref{dowmono}{Dowker,J.S., {\it Remarks on spherical monodromy defects  for free scalar fileds}, arXiv:2104.09419.}
      \ref{C+}{Chalabi,A., Herzog,C.P., O'Bannon,A., Robinson,B. and Sisti,J., {\it Weyl Anomalies of Four Dimensional Conformal Boundaries and Defects}, arXiv:2111.14713.}
      \ref{B+}{Bianchi, L., Chalabi, A., Proch\'azka, V., Robinson, B. and Sisti, J.,{\it Monodromy Defects in Free Field Theories}, arXiv:2104.01220.}
      \ref{Newman}{Newman, F. W. {\it On logarithmic integrals of the second order}, {\it Cambridge and Dublin Mathematical Journal}  {\bf 2} (1847)  77,172.}
     \ref{Dowstatistics}{Dowker,J.S. {\it Remarks on non--standard statistics}, \jpa{18} {1985} 3521.}
       \ref{Norlund}{N\"orlund,N.E. \am{43}{1922}{121}.}
    \ref{Norlund1}{N\"orlund,N.E. {\it Differenzenrechnung} (Springer--Verlag, 1924, Berlin.)}
 \ref{Dowcone}{Dowker,J.S. {\it Quantum field theory on a cone},\jpa{10}{1977}{115}.}
     \ref{Dowvector}{Dowker,J.S. {\it Lens space matter determinants in the vector model},
     arXiv:1405.7646.}
    \ref{BCPRS}{Bianchi,L.,Chalabi,A., Proch\'azka,V., Robinson,B. and Sisti,J., {\it Monodromy Defects in Free Field Theories}, arXiv:2104.01220.}
\ref{Dowtechnote}{Dowker,J.S. {\it A technical note on the calculation of GJMS (Rac and Di) operator determinants}, arXiv:1807.11872   }
     \ref{GHJK}{Giombi,S., Helfenberger,E., Ji,Z. and Khanchandani,H. {\it Monodromy Defects from Hyperbolic Space}, arXiv:2102.11815. }
     \ref{CCS}{Choi,J., Cho, Y-M and Srivastava, H.M, {\it Math.Scand.} {\bf105 } (2009) 199.}
     \ref{Dowchargedrenyi}{Dowker,J.S. {\it Charged R\'enyi entropies for free scalar fields}, \jpa{50} {2016} {165401},arXiv:1512.01135}
      \ref{CandB}{Crandall, R.E.  and Buhler, J.P. {\it On the evaluation of Euler sums}, {\it
   Experimental Math.} {\bf 3}  (1994) 275.}
     \ref{KandK}{Kurokawa,N. and Koyama, S-Y, {\it Multiple sine functions}, {\it Forum Mathematica}, {\bf15} (2003) 839.}
    \ref{KandK2}{Koyama, S-Y, and Kurokawa, N, {\it Euler's Integrals and Multiple sine functions}, \pams{133}{2004}{1257}.}
    \ref{Barnesa}{Barnes,E.W. {\it Trans. Camb. Phil. Soc.} {\bf 19} (1903)
  374.}
  \ref{DandD}{Diaz,D.E. and Dorn,H. {\it Partition functions and double trace
    deformations in AdS/CFT}, {\it JHEP} {\bf 0705} (2007) 46.}
     \ref{KPS2}{Klebanov,I.R., Pufu,S.S. and Safdi,B.R. {\it JHEP} {\bf 1110} (2011) 038.}
  \ref{Sch1}{Schr\"odinger, E.W. {\it Expanding Universes} (C.U.P. 1956. Cambridge).}
   \ref{Sch2}{Schr\"odinger, E.W. {\it Proc. Roy. Irish Acad.} {\bf 46A} (1946) 25.}
  \ref{Barnesb}{E.W.Barnes {\it Trans. Camb. Phil. Soc.} {\bf 19} (1903)
  426.}
    \ref{Wenger}{Wenger,D.L. \jmp{8}{1967}{135}.}
  \ref{Elizalde}{Elizalde,E. {\it Math. of Comp.} {\bf 47} (1986) 347.}
  \ref{CandT}{Copeland,E. and Toms,D.J. \np {255}{1985}{201}.}
    \ref{doweven}{Dowker,J.S. {\it Entanglement entropy on even spheres} arXiv:1009.3854.}
  \ref{Dowmasssphere}{Dowker,J.S. {\it Massive sphere determinants} arXiv:1404.0986.}
  \ref{dowtwistspin}{Dowker,J.S. {\it Conformal weights of charged Renyi entropy twist operators for free Dirac fields in arbitrary dimensions.} arXiv:1510.08378  .}
   \ref{dowtwist}{Dowker,J.S. {\it Conformal weights of charged R\'enyi entropy twist
   operators for free scalar fields in arbitrary dimensions}, \jpa{49}{2016}{145401}, arXiv:1509.00782.}
   \ref{dowsignch}{Dowker,J.S. \jpa{2}{1969}{267}.}
    \ref{Dowmultc}{Dowker,J.S. \jpa{5}{1972}{936}.}
   \ref{Dowrenexp}{Dowker,J.S. {\it Expansion of R\'enyi entropy for free scalar fields}
   arXiv:1408.0549.}
        \ref{Dowcascone}{Dowker,J.S. {\it Casimir effect around a cone}, \prD{36}{1987}{3095}.}
   \ref{EandH}{Elvang, H and  Hadjiantonis,M.  {\it Exact results for corner contributions to
   the entanglement entropy and R\'enyi entropies of free bosons and fermions in 3d} arXiv:
   1506.06729 .}
   \ref{CandH}{Casini H., and Huerta,M. \jpa{42}{2009}{504007}.}
   \ref{CaandH}{Casini,H. and Huerta,M. {\it Entanglement entropy on the n-sphere},\plb{694}{2010}{167}.}
    \ref{Dow7}{Dowker,J.S. \jpa{25}{1992}{2641}.}
    \ref{Dowcosecs}{Dowker,J.S. {\it On sums of powers of cosecs}, arXiv:1507.01848.}
    \ref{Jeffery}{Jeffery, H.M. \qjm{6}{1864}{82}.}
   \ref{BMW}{Bueno,P., Myers,R.C. and Witczak--Krempa,W. {\it Universal corner entanglement
   from twist operators} arXiv:1507.06997.}
  \ref{BMW2}{Bueno,P., Myers,R.C. and Witczak--Krempa,W. {\it Universality of corner
  entanglement in conformal field theories} arXiv:1505.04804.}
  \ref{BandM}{Bueno,P., Myers,R.C. {\it Universal entanglement for higher dimensional
  cones} arXiv:1508.00587.}
   \ref{B}{Belin,A.,Hung,L-Y., Maloney,A., Matsuura,S., Myers,R.C. and Sierens,T.\break
 {\it Holographic Charged R\'enyi Entropies}, {\it JHEP} {\bf 12} (2013) 059, arXiv:1310.4180.}
     \ref{Dowren}{Dowker,J.S. \jpamt {46}{2013}{2254}.}
     \ref{DandB}{Dowker,J.S. and Banach,R. {\it Quantum field theory on Clifford-Klein space--times. The effective Lagrangian and vacuum stress-energy tensor}, \jpa{11}{1978}{2255}.}
     \ref{Steffensen}{Steffensen,J.F. {\it Interpolation}, (Williams and Wilkins,
    Baltimore, 1927).}
    \ref{CaandH}{Casini,H. and Huerta,M. \plb{694}{2010}{167}.}
   \ref{Diaz}{Diaz,D.E. JHEP {\bf 7} (2008)103.}
     \ref{ChandD}{Chang,P. and Dowker,J.S. {\it Vacuum energy on orbifold factors of spheres}, \np{395}{1993}{407}.}
    \ref{DowGJMS}{Dowker,J.S. {\it Determinants and conformal anomalies of GJMS operators on spheres}, arXiv:1010.0566 ,\jpa{44}{2011}{115402}.}
    \ref{Doweven}{Dowker,J.S. {\it Entanglement entropy on even spheres.}
    arXiv:1009.3854.}
     \ref{Dowodd}{Dowker,J.S. {\it Entanglement entropy on odd spheres.}
     arXiv:1012.1548.}
    \ref{KPSS}{Klebanov,I.R., Pufu,S.S., Sachdev,S. and Safdi,B.R.
    {\it JHEP} 1204 (2012) 074.}
    \ref{Dowcmp}{Dowker,J.S. {\it Effective action in spherical domains}, arXiv:hep-th/9306154, \cmp{162}{1994}{633}.}
      \ref{Dowhyp}{Dowker,J.S. \jpa{43}{2010}{445402}.}
\end{putreferences}

\bye